\documentclass[fleqn,usenatbib]{mnras}
\usepackage{graphicx}	
\usepackage{amsmath}	
\usepackage{amssymb}	
\usepackage{multicol}        
\usepackage{bm}		
\usepackage{pdflscape}	
\usepackage[T1]{fontenc}
\usepackage{ae,aecompl}
\usepackage{newtxtext,newtxmath}
\usepackage{epsfig}
\usepackage{epstopdf}
\usepackage{booktabs}
\usepackage{footnote}
\usepackage{float}
\usepackage{multirow}
\usepackage{newtxtext,newtxmath}
\usepackage[T1]{fontenc}
\usepackage{graphicx}	
\usepackage{amsmath}
\usepackage{makecell,tabularx}
\DeclareRobustCommand{\VAN}[3]{#2}
\let\VANthebibliography\thebibliography
\def\thebibliography{\DeclareRobustCommand{\VAN}[3]{##3}\VANthebibliography}

\def\be{\begin{equation}}
\def\ee{\end{equation}}
\def\bea{\begin{eqnarray}}
\def\eea{\end{eqnarray}}

\title[Observational Constraints and Cosmological Implications of Scalar-Tensor $f(R, T)$ Gravity]{Observational Constraints and Cosmological Implications of Scalar-Tensor $f(R, T)$ Gravity}

\author[A. Bouali et al.]{
    Amine Bouali$^{1}$\thanks{E-mail: \href{mailto:a1.bouali@ump.ac.ma}{a1.bouali@ump.ac.ma}},
    Himanshu Chaudhary$^{2,3,4}$\thanks{E-mail: \href{mailto:himanshuch1729@gmail.com}{himanshuch1729@gmail.com}},
    Tiberiu Harko$^{5,6,7}$\thanks{E-mail: \href{mailto:t.harko@ucl.ac.uk}{t.harko@ucl.ac.uk}},
    Francisco S. N. Lobo$^{8,9}$\thanks{E-mail: \href{mailto:fslobo@fc.ul.pt}{fslobo@fc.ul.pt}},
    Taoufik Ouali$^{1}$\thanks{E-mail: \href{mailto:t.ouali@ump.ac.ma}{t.ouali@ump.ac.ma}}, \newauthor
    and Miguel A. S. Pinto$^{8,9}$\thanks{E-mail: \href{mailto:mapinto@fc.ul.pt}{mapinto@fc.ul.pt}}
    \\
    $^{1}$Laboratory of Physics of Matter and Radiation, Mohammed I University, BP 717, Oujda, Morocco\\
    $^{2}$Department of Applied Mathematics, Delhi Technological University, Delhi-110042, India\\
    $^{3}$Pacif Institute of Cosmology and Selfology (PICS) Sagara, Sambalpur 768224, Odisha, India,\\
    $^{4}$Department of Mathematics, Shyamlal College, University of Delhi, Delhi-110032, India.\\
    $^{5}$Department of Physics, Babes-Bolyai University, Kogalniceanu Street, Cluj-Napoca 400084, Romania\\
    $^{6}$Department of Theoretical Physics, National Institute of Physics and Nuclear Engineering (IFIN-HH), Bucharest, 077125 Romania\\
    $^{7}$Astronomical Observatory, 19 Ciresilor Street, Cluj-Napoca 400487, Romania\\
    $^{8}$Instituto de Astrofísica e Ciências do Espaço, Faculdade de Ciências da Universidade de Lisboa, Edifício C8, Campo Grande, P-1749-016\\
    Lisbon, Portugal\\
    $^{9}$Departamento de F\'{i}sica, Faculdade de Ci\^{e}ncias da Universidade de Lisboa, Edifício C8, Campo Grande, P-1749-016 Lisbon, Portugal
}

\pubyear{2023}
\begin{document}
\label{firstpage}
\pagerange{\pageref{firstpage}--\pageref{lastpage}}
\maketitle

\begin{abstract}
Recently, the scalar-tensor representation of $f (R,T)$ gravity was used to explore gravitationally induced particle production/annihilation. Using the framework of irreversible thermodynamics of open systems in the presence of matter creation/annihilation, the physical and cosmological consequences of this setup were investigated in detail. In this paper, we test observationally the scalar-tensor representation of $f(R,T)$ gravity in the context of the aforementioned framework, using the Hubble and Pantheon+ measurements. The best fit parameters are obtained by solving numerically the modified Friedmann equations of two distinct cosmological models in scalar-tensor $f(R, T)$ gravity, corresponding to two different choices of the potential, and by performing a Markov Chain Monte Carlo analysis. The best parameters are used to compute the cosmographic parameters, i.e., the deceleration, the jerk and the snap parameters. Using the output resulting from the Markov Chain Monte Carlo analysis, the cosmological evolution of the creation pressure and of the matter creation rates are presented for both models. To figure out the statistical significance of the studied scalar-tensor $f(R,T)$ gravity, the Bayesian and the corrected Akaike information criteria are used. The latter indicates that the first considered model in scalar-tensor $f(R,T)$ gravity is statistically better than $\Lambda$CDM, i.e., it is more favored by observations. Besides, a continuous particle creation process is present in  Model 1. On the other hand, for large redshifts, in  Model 2 the particle creation rate may become negative, thus indicating the presence of particle annihilation processes. However, both models lead to an accelerating expansion of the Universe at late times, with a deceleration parameter equivalent to that of the $\Lambda$CDM model.
\end{abstract}

\begin{keywords}
Observations -- Dark Energy -- Methods: Statistical -- Methods: Numerical -- Methods: Data Analysis -- Miscellaneous

\end{keywords}

\section{Introduction}
General Relativity (GR) \citep{1} can be unequivocally considered as one of the most important physical theories developed in recent times, and it is the standard theoretical foundation for the modern comprehension of gravitational phenomena. Essentially, GR is a geometric theory of gravity, in which the gravitational interaction is explained by assuming a Riemannian structure of the spacetime. The recent detection of the gravitational waves by the LIGO (Laser Interferometer Gravitational-Wave Observatory) and Virgo collaboration \citep{2}, which started a new and important era for the astronomical, astrophysical, and cosmological probes of the Universe, confirmed once again the major predictive power of this theory.
However, despite its remarkable successes and achievements, GR faces several problems on both small and large scales. On the former, we have the presence of singularities in black hole solutions, like, for example, in the Schwarzschild solution \citep{3}. The appearance of singularities is unavoidable in GR \citep{4,5}. We must also take into account the fact that the theory is not renormalizable at high energies, and therefore cannot be quantized \citep{6}. Moreover, while all the standard model interactions are conformally invariant at the classical level, GR is not \citep{7,8,9}. Hence, from the perspective of GR, the gravitational interaction cannot be understood, and treated in the same manner as the remaining fundamental interactions, which leads to fundamental problems in the quantization of the theory.

More recently, based on observations of Supernovae Type IA, it was found that the Universe is presently undergoing an accelerated expansion that is thought to be caused by an exotic fluid, dubbed as dark energy \citep{11,12}. Einstein's cosmological constant $\Lambda$ \citep{10} is a particular type of this fluid, whose physical nature is still yet to be understood. In fact, the cosmological constant is a subject of great controversy, since if we assume that it is the vacuum energy, then the theoretical value predicted by quantum field theory is largely (of an order of $10^{120}$) different from the observational value obtained from cosmological analysis, which, in short,  constitutes the famous Cosmological Constant Problem \citep{13,14,15}. Furthermore, GR is unable to describe correctly the galactic rotation curves \citep{16,18,19}, without assuming the presence of a non-luminous matter, called as dark matter, to fill the mass gap that is supposedly missing. Dark matter is assumed to interact only gravitationally with normal baryonic matter. The variety of theoretical models for this kind of matter is very rich, with the main candidates being WIMPs (Weakly Interacting Massive Particles), sterile neutrinos, axions, and supersymmetric particles (for reviews of the many proposed particles that could be dark matter see  \citep{20,21,22,23}. Still, so far, no terrestrial experiment has directly detected dark matter particles, or any of their possible signatures (for a recent review on the state of the art regarding dark matter searches see \citet{24}.

 Recently, the $\Lambda$CDM model must also face a number of other important challenges, resulting from the significant increase of the precision of the cosmological observations. One of the most challenging present day cosmological problems is the important difference between the expansion rate of the Universe as determined from the Cosmic Microwave Background  (CMB) experiments and the low redshift (local) measurements. These important differences in the determinations of the present-day value of the Hubble constant led to what is generically called the Hubble tension.  For a detailed discussion of the Hubble tension, and of its possible solutions see \citep{T1} and references therein.  The Hubble constant ($\textrm{H}_0$) as measured by the Planck satellite has the value of $66.93 \pm 0.62$ km/ s/ Mpc, while the SHOES collaboration obtained a value of $73.24\pm 1.74$ km/ s/ Mpc \citep{T1}. If the Hubble tension is real, it strongly points toward the necessity of investigating new gravitational theories, and of the replacement the $\Lambda$CDM model with alternative cosmological models.

A possible path in an attempt to solve the aforementioned problems is to modify the Einstein-Hilbert action. A common example is the $f(R)$ modified theory of gravity \citep{25,26}, in which one substitutes the Ricci scalar, $R$, by an arbitrary and well-behaved function of the same quantity, $f(R)$. Such an extension allows for a broader and more diverse gravitational dynamics at low energies, which can, for example, explain the late-time cosmic acceleration without the necessity of considering the existence of a cosmological constant \citep{27}. Indeed, admitting that the nature of the gravitational interaction changes on cosmological scales could pave the way to new breakthroughs in cosmology \citep{28,29}.

A particularly interesting class of modified theories of gravity are represented by those that contain nonminimal geometry-matter couplings, i.e., (direct) interaction terms between geometry (describing gravitational fields), and matter (non-gravitational fields) in the corresponding Lagrangian density \citep{30,31,32,33,34,35,36,37,38,39,40,41,42,43,45}.  In particular, in these types of theories of gravity, the matter energy-momentum tensor is not conserved generally, and an extra-force is generated due to the presence of the geometry-matter coupling \citep{30,31}. The most natural physical interpretation of the matter non-conservation is by assuming the presence of particle creation, which implies the irreversible matter/energy transfer from gravity (curvature) to matter fields.  Hence, in models with geometry-matter couplings, the natural description of the thermodynamic phenomena occurs within the thermodynamics of open systems in the presence of irreversible processes. However, we would like to point out that presently the existence of the active matter creation processes in the Universe represents a hypothesis, which is not (yet) supported by any direct observational or experimental evidence.

A prominent example of a theory with a non-minimal curvature-matter coupling is the $f(R,T)$ gravity theory \citep{46}, where $T$ is the trace of the energy-momentum tensor $T_{\mu\nu}$. As the name suggests, in this case, the Einstein-Hilbert action is generalized by the substitution of the Ricci scalar by a general $f(R,T)$ function. $f(R,T)$ gravity can be also seen as an extension of the $f(R)$ theory, in which now it is possible to have a direct exchange of energy and momentum between curvature and matter. The exploration of this theory in the usual geometrical representation was extensively done in the literature (see, for instance,  \citep{47,48,49,50}.

Additionally, the recently proposed scalar-tensor representation for this theory in \citep{51} has triggered an intensive research effort \citep{52,53,54,55}. The scalar-tensor representation not only helps in decreasing the complexity of the mathematical formalism by reducing the order of the metric in the field equations, but also allows to explore cosmological solutions by using a dynamical system approach. Moreover, the scalar-tensor $f(R,T)$ gravity has been recently considered in the context of irreversible matter creation, demonstrating capabilities of explaining the entropy production alongside a creation of cosmological matter, which can trigger the recent cosmological accelerated expansion of the Universe \citep{56}.
Although it is very important to formulate new modified theories of gravity, in order to reach a more complete description of the gravitational phenomena, it is equally important to test and constrain the existing theories as well, by using the most recent, and the best available observational probes, such as, for example, the Hubble measurements and Supernova type Ia (i.e Pantheon+) data.

Given that, in this paper, we have performed a detailed comparison with the aforementioned cosmological observations of two particular cosmological models in the scalar-tensor representation of  $f(R,T)$ gravity, in which the gravitational interaction is described by two scalar  fields $\varphi$ and $\psi$, and a self-interaction potential $U(\varphi, \psi)$. The cosmological models adopted in the present study correspond to two different choices of the interaction potential of the theory, which is assumed to take the forms $U(\varphi,\psi)=\alpha \varphi +\beta \varphi ^2-(1/2\gamma)\psi^2$, and $U(\varphi,\psi)=\alpha \varphi ^n\psi^{-m}+\beta \varphi$, respectively, with the first potential having a purely additive structure, while the second potential has an additive-multiplicative structure in the two basic fields $\varphi$ and $\psi$. Some general features of the cosmologies corresponding to these choices have been already investigated in \citep{56}.   In \citep{56}, the particle creation rates, the creation pressure, the temperature evolution, and the comoving entropy have been obtained by using systematic covariant thermodynamic formulation. The thermodynamics of open systems was considered together with the field equations of the scalar-tensor representation of the $f(R,T)$ theory, and thus an extension of the  $\Lambda$CDM paradigm was obtained. One of the important features of this approach is that the matter creation rates, as well as the creation  pressure, are included in the energy-momentum tensor of the cosmological fluid. Specific cosmological models were also constructed by using the scalar-tensor $f (R ,T )$ gravity theory, and a comparison with the $\Lambda$CDM model, as well as  the observational data for the Hubble function was performed by using the trial and error approach for fixing the initial conditions of the generalized Friedmann equations,  and of the model parameters. Moreover, the thermodynamic properties of the cosmological models in the presence of particle creation were investigated in both the high and the low redshift regimes.

It is important to point out that the cosmological models depend on several arbitrary constants (parameters) $\alpha$, $\beta$ and $\gamma$, whose values cannot be predicted theoretically, but must be obtained from a comparison with the observational data.

 As a first step in our analysis of the $f(R,T)$ cosmological models, we have solved numerically the Friedmann equations in the scalar-tensor representation of the $f(R,T)$ gravity for the two given potentials.  Then, by means of a Markov Chain Monte Carlo (MCMC) analysis, we have extracted the best fit values of the parameters of the two models $\alpha$, $\beta$, and $\gamma$, and we have also obtained their mean values.
 Based on these statistical results, the two $f(R, T)$ gravity models were compared, analyzed, and classified with respect to the standard $\Lambda$CDM model by using the  Bayesian and the corrected  Akaike Information Criterion.

 As a second step of this work,  we have analyzed in detail the footprints of these fitting results on the current acceleration of the Universe by studying the evolution of the cosmographic parameters. Indeed, from the observational data, the cosmographic parameters can describe the past history and the evolution of the scale factor through a series of its derivatives.  Furthermore, as they are model-independent tools, the kinematic parameters are used to discriminate between competing cosmological models describing the current cosmic speed up without assuming any cosmological model \citep{57,58}.  Cosmography has already been subject of many studies in modified gravity such as in  $f(R)$ gravity \citep{59,60,61,62,63,64}, in  $f(T)$ teleparallel gravity \citep{65,66}, in $f(R,T)$ gravity theory \citep{67}, and, recently, in the Barthel-Kropina dark energy model \citep{68}.
 Another powerful tool considered in this paper and recently used to discriminate between alternative dynamical dark energy models from $\Lambda$CDM is the $Om(z)$ diagnostic \citep{69,70}. We have also investigated the $Om(z)$ diagnostic for the considered $f(R,T)$ models, and compared its predictions with the $\Lambda$CDM standard cosmology.

This paper is organized in the following manner. The modified field equations of $f(R,T)$ gravity in both geometrical and scalar-tensor representations are presented in Section \ref{sec:fRTintro2}. The general cosmological assumptions and equations are considered in Section \ref{sec:cosmo}. We consider two different functional forms for the scalar interaction potential, which specify the two particular cosmological model we are working with. The generalized Friedmann equations for two distinct models are also explicitly written down. In Section \ref{cosmtests}, we perform several cosmological tests to constrain the free parameters of the considered particular cosmological models. Finally, we discuss and conclude the results obtained in our paper in Section \ref{conclusions}.

\section{Field equations of $f\left(R,T\right)$ gravity, and their cosmological implications}\label{sec:fRTintro}

In the present Section, we briefly review the fundamentals of the $f(R,T)$ modified theory of gravity, and of its scalar-tensor representation. The (non-)conservation equations of the matter energy-momentum tensor in these representations are also presented.  In order to consider the observational tests of the theory we will consider the cosmological implications of the scalar-tensor representation of the $f(R,T)$, by assuming an isotropic, homogeneous and spatial flat geometry. The generalized Friedmann equations, describing the large scale evolution are introduced first in a general form. Two specific cosmological models, corresponding to two different choices of the self-interaction potential $U(\varphi,\psi)$ are also presented. These models will be used in the next Section as toy models for the cosmological tests of the theory.

\subsection{Metric and scalar tensor representation of $f(R,T)$ gravity}\label{sec:fRTintro2}

In the present Subsection, we write down the basic relations in the two equivalent representations of the $f(R,T)$ gravity theory. Since we are working under the metric formalism, henceforth we denote the geometrical representation as the metric representation.

\subsubsection{Metric representation}

The action for the $f\left(R,T\right)$ gravity theory is postulated to be \citep{46}
\begin{equation}\label{eq:fRTaction-original}
    S = \frac{1}{2\kappa^2} \int_{\Omega}\sqrt{-g} \, f(R,T) d^4 x+ \int_{\Omega} \sqrt{-g} \, \mathcal{L}_m d^4 x,
\end{equation}
where we have denoted $\kappa^2=8\pi G/c^4$, with $G$ the gravitational constant, and $c$ is the speed of light in vacuum. Moreover, $g$ is the determinant of the metric tensor $g_{\mu\nu}$, while  $\Omega$ denotes the 4-dimensional spacetime Lorentzian manifold on which the gravitational field acts, and on which it is possible to define a set of coordinates $\{x^\mu\}$. The gravitational Lagrangian density is given by $f\left(R,T\right)$, where $f$ is an arbitrary function of the Ricci scalar $R=g^{\mu\nu}R_{\mu\nu}$, with $R_{\mu\nu}$ denoting the Ricci tensor, and $T$ is the trace $T=g^{\mu\nu}T_{\mu\nu}$ of the matter energy-momentum tensor $T_{\mu\nu}$. The matter energy-momentum tensor is defined in terms of the variation of the matter Lagrangian density $\mathcal L_m$ as
\begin{equation}
T_{\mu\nu}=-\frac{2}{\sqrt{-g}}\frac{\delta\left(\sqrt{-g}\mathcal L_m\right)}{\delta g^{\mu\nu}}.
\end{equation}
Henceforth, we adopt a system of geometrized units in such a way that $G=c=1$, and thus $\kappa^2=8\pi$.

By applying the action variational principle, $\delta S=0$, the variation of Eq.~\eqref{eq:fRTaction-original} with respect to the metric tensor $g_{\mu\nu}$ yields the modified gravitational field equations see \citep{46} for a detailed derivation)
\bea\label{eq:fields-original}
    f_R R_{\mu\nu}-\frac{1}{2}g_{\mu\nu}f(R,T) &+& \left(g_{\mu\nu}\square-\nabla_\mu\nabla_\nu\right)f_R \nonumber\\
    &=& \kappa^2 T_{\mu\nu}-f_T (T_{\mu\nu}+\Theta_{\mu\nu}),
\eea
where the subscripts $R$ and $T$ denote the partial derivatives of $f$ with respect to the respective quantities, respectively, $\nabla_\mu$ is the covariant derivative and $\square\equiv\nabla^\sigma\nabla_\sigma$ denotes the D’Alembert operator in its standard form. Furthermore, the auxiliary tensor $\Theta_{\mu\nu}$, introduced in Eq. \eqref{eq:fields-original}, is defined as
\begin{equation}\label{eq:Theta-varT}
    \Theta_{\mu\nu}\equiv g^{\rho\sigma}\frac{\delta T_{\rho\sigma}}{\delta g^{\mu\nu}}.
\end{equation}

The conservation equation for $f\left(R,T\right)$ gravity can be obtained by taking the divergence of Eq.~\eqref{eq:fields-original} and using the identity $\left(\square\nabla_\nu-\nabla_\nu\square\right)f_R=R_{\mu\nu}\nabla^\mu f_R$ \citep{30}. By doing so, we obtain
\bea\label{eq:conserv-general}
  &&  (\kappa^2-f_T)\nabla^\mu T_{\mu\nu}=\left(T_{\mu\nu}+\Theta_{\mu\nu}\right)\nabla^\mu f_T \nonumber\\
    &&+f_T\nabla^\mu\Theta_{\mu\nu}+f_R \nabla^\mu R_{\mu\nu}-\frac{1}{2}g_{\mu\nu}\nabla^\mu f.
\eea

By looking at the form of Eq. \eqref{eq:conserv-general}, it is evident that the matter energy-momentum tensor is not necessarily conserved. In fact, one should note that, in general, modified theories of gravity that may possess non-minimal geometry-matter couplings \citep{30,31,32,33,34,35,36,37,38,39,40,41,42,43,44,45} imply the possible non-conservation of the matter energy-momentum tensor, with $\nabla_\mu T^{\mu\nu} \neq 0$. This result is standardly interpreted as a transfer of energy from the geometry sector to the matter sector \citep{42,43,44,45}.

Next, we consider the equivalent scalar-tensor representation of $f(R, T)$ gravity, which will be utilized henceforth.

\subsubsection{Scalar-Tensor Representation}\label{subsec:scalar-tensor}

One of the interesting and important properties of modified theories of gravity, featuring extra scalar degrees of freedom in comparison to GR, is that one can deduce a dynamically equivalent scalar-tensor theory.  As such, it is possible to construct an equivalent scalar-tensor action to Eq. \eqref{eq:fRTaction-original}, in which two extra scalar fields are added as fundamental fields, each one related to a degree of freedom of the metric representation of $f(R, T)$ gravity \citep{51}. In the particular case of this theory, the degrees of freedom are $R$ and $T$.

To obtain the scalar-tensor representation, one introduces the first two auxiliary fields, $\eta$ and $\zeta$, respectively. Then one rewrites Eq. \eqref{eq:fRTaction-original} in the following form
\begin{eqnarray}\label{eq:STaction-intro}
    S &=& \frac{1}{2\kappa^2} \int_{\Omega}\sqrt{-g} \big[f(\eta,\zeta)+ (R-\eta)f_\eta
		\nonumber \\
    && +(T-\zeta)f_\zeta\big] d^4 x  + \int_{\Omega} \sqrt{-g}\mathcal{L}_m  d^4 x ,
\end{eqnarray}
where the subscripts $\eta$ and $\zeta$ indicate the operation of partial derivation with respect to the considered variables. After taking the variation of the action \eqref{eq:STaction-intro} with respect to the fields $\eta$ and $\zeta$, we find that the equations of motion in these variables are
\begin{eqnarray}
    (R-\eta)f_{\eta\eta}+(T-\zeta)f_{\eta\zeta} &=& 0, \label{eq:motion-alpha} \\
    (R-\eta)f_{\zeta\eta}+(T-\zeta)f_{\zeta\zeta} &=& 0. \label{eq:motion-beta}
\end{eqnarray}
Now, we rewrite the system of Eqs. \eqref{eq:motion-alpha} and \eqref{eq:motion-beta} in a matrix form,  $\mathcal M\textbf{u}=0$ as
\begin{equation}\label{9}
\begin{pmatrix}
f_{\eta\eta} & f_{\eta\zeta} \\
f_{\zeta\eta} & f_{\zeta\zeta}
\end{pmatrix}
\begin{pmatrix}
R-\eta\\
T-\zeta
\end{pmatrix}=0.
\end{equation}

A unique solution of the above algebraic equation is obtained if and only if det $\mathcal M \neq 0$, a condition which gives the relation  $f_{\eta\eta}f_{\zeta\zeta}\neq f_{\eta\zeta}^2$, which must be satisfied by the function $f(\eta, \zeta)$. Hence, it follows that Eq. (\ref{9}) has the unique solution given by $R=\eta$ and $T=\zeta$. After inserting these results back into Eq.~\eqref{eq:STaction-intro}, one easily verifies that this equation takes the form of the initial action \eqref{eq:fRTaction-original}. Thus, we have proved the equivalence between the two representations, and we have shown that the scalar-tensor representation is well-defined and mathematically consistent.

As a last step, we introduce two scalar fields, $\varphi$ and $\psi$, as well as an interaction potential between them, $U\left(\varphi,\psi\right)$, according to the definitions
 \begin{equation}\label{eq:varphi&psi}
     \varphi\equiv\frac{\partial f}{\partial R} ,\qquad
    \psi\equiv\frac{\partial f}{\partial T},
 \end{equation}
 and
\begin{equation}\label{eq:potential}
    U(\varphi,\psi) \equiv -f(\eta,\zeta)+ \eta \varphi  + \zeta \psi ,
\end{equation}
respectively. Now we rewrite Eq. \eqref{eq:STaction-intro}, and display the action of the equivalent scalar-tensor representation of $f(R,T)$ gravity as
\bea\label{eq:STaction}
    S &=& \frac{1}{2\kappa^2} \int_{\Omega} \sqrt{-g} \left[\varphi R+\psi T - U(\varphi, \psi)\right]d^4 x \nonumber\\
    &&+ \int_{\Omega} \sqrt{-g} \mathcal{L}_m d^4 x .
    \eea

The scalar field $\varphi$ is the analogue of a Brans-Dicke type scalar field with parameter $\omega_{BD}=0$, and with a self-interaction potential $U$ \citep{Far}. Additionally to the scalar field $\varphi$, in  $f(R,T)$ gravity we have a second degree of freedom, corresponding to the arbitrary dependence of the action on $T$. This second degree of freedom is also related to another scalar field, $\psi$.

Accordingly to what was previously stated, the action \eqref{eq:STaction} depends now on three independent fields, the metric tensor $g_{\mu\nu}$ and the two scalar fields $\varphi$ and $\psi$. By varying the action Eq. \eqref{eq:STaction} with respect to $g_{\mu\nu}$ yields the field equations
\begin{eqnarray}\label{eq:fields}
\varphi R_{\mu\nu}&-&\frac{1}{2}g_{\mu\nu}\left(\varphi R + \psi T - U\right)-(\nabla_\mu\nabla_\nu-g_{\mu\nu}\square)\varphi \nonumber\\
&=& \kappa^2 T_{\mu\nu} -\psi (T_{\mu\nu} + \Theta_{\mu\nu}).
      \end{eqnarray}

It is also possible to derive the above field equation (\ref{eq:fields}) directly by using Eq.~\eqref{eq:fields-original} via the substitution of the quantities  defined in Eqs.~\eqref{eq:varphi&psi} and \eqref{eq:potential}, respectively, by taking $\eta=R$ and $\zeta=T$. Furthermore, the variation of Eq.~\eqref{eq:STaction} with respect to the scalar fields $\varphi$ and $\psi$ gives,
\begin{equation}\label{eq:Vphi}
    U_{\varphi} = R, \qquad   U_{\psi} = T.
\end{equation}

In addition to this, by using Eqs.~(\ref{eq:varphi&psi}) and (\ref{eq:potential}), with $\eta=R$ and $\zeta=T$, and the mathematical identity $\nabla^\mu\left[R_{\mu\nu}-(1/2)g_{\mu\nu}R\right]=0$, we obtain the results that the conservation equation for $f\left(R,T\right)$ gravity in the scalar-tensor representation takes the form
\bea\label{eq:conserv-general2}
  \nabla^\mu T_{\mu\nu}&=&\frac{\left(T_{\mu\nu}+\Theta_{\mu\nu}\right)\nabla^\mu\psi}{(\kappa^2-\psi)}+\frac{\psi\nabla^\mu\Theta_{\mu\nu}}{(\kappa^2-\psi)} \nonumber\\    &&-\frac{1}{2}g_{\mu\nu}\frac{R\nabla^\mu\varphi+\nabla^\mu\left(\psi T-U\right)}{(\kappa^2-\psi)}.
\eea

\subsection{Cosmological evolution, and cosmological models}\label{sec:cosmo}


We assume that the metric of the Universe is given by the flat, homogeneous and isotropic Friedmann-Lemaitre-Robertson-Walker (FLRW) expression, which in spherical coordinates $(t,r,\theta,\phi)$ is given by
\begin{equation}\label{eq:FLRW-metric}
    ds^2 = -dt^2+a^2(t)\left[dr^2+r^2\left(d\theta^2+\sin^2\theta d\phi^2\right)\right],
\end{equation}
where $a(t)$ is the scale factor. Moreover, we introduce the Hubble function defined as $H=\dot{a}/a$, where, in the following, an overdot denotes the derivative with respect to the cosmological time $t$.

We also conjecture that the matter content of the Universe can be represented as an isotropic perfect fluid, with the energy-momentum tensor of the matter $T_{\mu\nu}$ given by
\begin{equation}\label{eq:em-fluid}
    T_{\mu\nu}=(\rho+p)u_\mu u_\nu +p g_{\mu\nu},
\end{equation}
where $\rho$ is the energy density of the cosmic matter, and $p$ is its pressure. The four-velocity $u^\mu$ of the cosmological fluid is normalized according to $u_\mu u^\mu=-1$. As for the matter Lagrangian density we adopt the expression $\mathcal L_m=p$ \citep{32}.  Our assumption of the isotropy of the matter distribution also implies that on average, and on large cosmological scales,  the matter creation process are isotropic. However, particle production phenomena, quantum or classical,  may not be necessarily isotropic. Anisotropic matter creation  naturally generates inhomogeneities at small cosmological or galactic scales, and leads to the presence of anisotropic stresses, which could play an important, or event determinant role in the structure formation. However, on large cosmological scales, after averaging the matter density over large cosmological volumes,  the cosmological dynamics can be described by the usual Friedmann-Lemaître-Robertson-Walker cosmology, in which the matter dynamics decouples completely from that describing particle creation inhomogeneities. Hence, in the following we consider that the possible physical or astrophysical properties inherited by the matter density from a possible anisotropic particle generation process, in the large scale phase, the average cosmological dynamics can be well described as taking place in a homogeneous and isotropic Universe. This assumption is also supported by the comparison of our model with the observational data, indicating that the particle creation model fits well the distance modulus observational data of type Ia supernovae, and thus it correctly describes the observed accelerated phase of the Universe.

Thus,  the auxiliary tensor $\Theta_{\mu\nu}$ becomes
\begin{equation}\label{eq:Theta-fluid}
    \Theta_{\mu\nu}=-2T_{\mu\nu}+p g_{\mu\nu}.
\end{equation}

In a cosmological setting all geometrical and thermodynamical quantities must depend only on the cosmological time $t$. Consequently, in the present approach,  $\rho=\rho\left(t\right)$, $p=p\left(t\right)$, $\varphi=\varphi\left(t\right)$, and $\psi=\psi\left(t\right)$, respectively.

Accordingly, the system of the cosmological field equations of the scalar-tensor representation of $f(R,T)$ gravity, representing the generalized Friedmann equations, takes the form \citep{56}
\be\label{Fr1}
3H^2=8\pi \frac{\rho}{\varphi}+\frac{3\psi}{2\varphi}\left(\rho-\frac{1}{3}p\right)+\frac{1}{2}\frac{U}{\varphi}-3H\frac{\dot{\varphi}}{\varphi},
\ee
\be\label{Fr2}
2\dot{H}+3H^2=-8\pi \frac{p}{\varphi}+\frac{\psi}{2\varphi}\left(\rho-3p\right)+\frac{1}{2}\frac{U}{\varphi}-\frac{\ddot{\varphi}}{\varphi}-2H\frac{\dot{\varphi}}{\varphi},
\ee
\be\label{Fr3}
U_{\varphi}=6\left(\dot{H}+2H^2\right), U_{\psi}=3p-\rho,
\ee
\bea\label{Fr4}
\hspace{-0.3cm}\dot{\rho}+3H(\rho +p)&=&\frac{3}{8\pi}\Bigg\{-\frac{\dot{\psi}}{2}\left(\rho-\frac{p}{3}+\frac{U_{\psi}}{3}\right)\nonumber\\
\hspace{-0.3cm}&&-\psi\left[H(\rho+p)+\frac{1}{2}\left(\dot{\rho}-\frac{1}{3}\dot{p}\right)\right]\Bigg\}.
\eea

To indicate the accelerating/decelerating nature of the cosmological expansion we will use the deceleration parameter $q$, defined as
\be
q=\frac{d}{dt}\frac{1}{H}-1=-\frac{\dot{H}}{H^2}-1.
\ee

From Eqs. (\ref{Fr1}) and (\ref{Fr2}), we find the following expression for the deceleration parameter
\be
q=\frac{1}{2}+\frac{3\left[ 4\pi \frac{p}{\varphi }-\frac{\psi }{4\varphi }%
\left( \rho -3p\right) -\frac{V}{4\varphi }+\frac{\ddot{\varphi}}{2\varphi }%
+H\frac{\dot{\varphi}}{\varphi }\right] }{8\pi \frac{\rho }{\varphi }+\frac{%
\psi }{2\varphi }\left( \rho -3p\right) +\frac{V}{2\varphi }-3H\frac{\dot{%
\varphi}}{\varphi }}.
\ee

In order to facilitate the comparison with the observational data, we will study the cosmological evolution in the redshift space, with the redshift coordinate $z$ defined as
\be
1+z=\frac{1}{a}.
\ee

In the following we will also adopt the thermodynamic interpretation of $f(R,T)$ gravity, as introduced in \citep{56}. Particle creation can be
formulated in a thermodynamic formalism in which one introduces the creation pressure $p_c$, given by \citep{56}
\be
p_c=\frac{\rho+p}{3H}\frac{\psi}{8\pi+\psi}\left(\frac{d}{dt}\ln \psi+\frac{1}{2}\frac{\dot{\rho}-\dot{p}}{\rho+p}\right),
\ee
which is related to the particle creation rate $\Gamma$ through the relation \citep{LIMA}
\be
p_c=-\frac{\rho+p}{3H}\Gamma.
\ee
Hence, for the particle creation rate we obtain the expression
\be
\Gamma=-\frac{\psi}{8\pi+\psi}\left(\frac{d}{dt}\ln \psi+\frac{1}{2}\frac{\dot{\rho}-\dot{p}}{\rho+p}\right).
\ee
 The expressions of the particle number creation rate $\Gamma$ and of the creation pressure $p_c$  follow directly from Eq.\eqref{eq:conserv-general2}, expressing the non-conservation of the matter energy-momentum tensor. The present model does admit de Sitter type, exponentially expanding solutions. Two such classes of solutions can be obtained, under the assumption of a constant matter density, or, alternatively, in the presence of a time varying matter density. By assuming $H=H_0={\rm constant}$ and $\rho=\rho_0={\rm constant}$, Eq.~(\ref{Fr2}) gives for the potential the expression
\be\label{pota}
U(\phi,\psi)=12H_0^2\varphi-\rho_0\psi +\Lambda_0,
\ee
where $\Lambda_0$ is an arbitrary integration constant. With this form of the potential, describing de Sitter type expansion, the general solution of the field equations was found in \citep{56} and corresponds to a cosmological dynamics during which both the particle creation rate and the creation pressure decrease exponentially in time. 

A second de Sitter type solution can be obtained for dust matter by assuming that the potential has the form \citep{56}
\be\label{potb}
U(\phi,\psi)=12H_0^2\varphi -\frac{1}{2\beta_1}\psi^2,
\ee
where $\beta _1$ is a constant. The derivative of the potential with respect to $\psi$ gives immediately $U_\psi=-\psi/\beta _1$, while from Eq.~(\ref{Fr2}) for $p=0$ we obtain $U_\psi=-\rho$, thus obtaining $\psi =\beta _1\rho$. For this choice of the potential the matter density decreases exponentially, while the creation rate becomes either a constant, for $\beta _1\rho>>8\pi$, $\Gamma \approx (9/5)H_0$, or decreases exponentially in the opposite limit as $\Gamma (t)\propto e^{-3H_0t}$.

In the following we write down the generalized Friedmann equations for two distinct cosmological models, which we will use as toy models for testing the validity of $f(R,T)$ gravity on cosmological scales.   The considered models correspond to choices of the potential that extend the de Sitter type forms of the potential (\ref{pota}) and (\ref{potb}) beyond the strict exponential phase, but still describe an accelerating Universe.  From a physical point of view, for a Universe with time varying matter density, the potential $\psi$ can be interpreted as an effective matter density.  As for $\varphi$, as one can see from Eq.~(\ref{Fr2}), it is fully determined by the expansion rate of the Universe, through the derivative of the potential. Hence, from a physical point of view, in a general sense, the potential describes the combined effects of the cosmological expansion rate, and of the matter density on the dynamical evolution of the Universe.

\subsubsection{Model 1: $U(\varphi,\psi)=\alpha \varphi +\beta \varphi ^2-\left(1/2\gamma \right) \psi ^2$}

As a first cosmological model in the scalar-tensor representation of $f(R,T)$
gravity we consider that the self-interacting potential $U$ has a simple
quadratic additive quadratic form, with
\begin{equation}
U(\varphi,\psi)=\alpha \varphi
+\beta \varphi ^2-\frac{1}{2\gamma} \psi ^2,
\end{equation}
where $\alpha$, $\beta$ and $\gamma$ are arbitrary constants to be determined from the observations.
For a dust Universe with $p=0$ we have $\psi =\gamma \rho$.  This form of the potential generalizes the expression (\ref{potb}) of $U(\varphi,\psi)$ by adding to it an additive quadratic correction term in $\varphi$. The deviation of the potential from the de Sitter form leads to an accelerating Universe, with non-exponential scale factor, and with time varying matter density,

After introducing a set of dimensionless and rescaled variables $\left(
h,\tau ,r,\xi ,\sigma ,\epsilon \right) $, defined as
\begin{eqnarray}
H &=&H_{0}h,\qquad \tau =H_{0}t,\qquad \rho =\rho _{c}r,  \nonumber \\
\alpha  &=&16\pi \rho _{c}\xi , \qquad \beta =8\pi \rho _{c}\sigma , \qquad \gamma =\frac{%
8\pi }{\rho _{c}}\epsilon ,
\end{eqnarray}%
where $H_{0}$ is the present-day value of the Hubble function, and $\rho
_{c}=3H_{0}^{2}/8\pi $, we obtain the generalized Friedmann equations of
Model 1 in the redshift space as given by \citep{56}
\begin{equation}
-\frac{dh(z)}{dz}=\frac{\xi }{(1+z)h(z)}+\frac{\sigma \varphi (z)}{(1+z)h(z)}%
-2\frac{h(z)}{(1+z)},  \label{m1}
\end{equation}%
\begin{equation}
-\frac{dr(z)}{dz}=-3\frac{r(z)(1+\epsilon r(z))}{(1+z)\left( 1+(5\epsilon
/2)r(z)\right) },  \label{m2}
\end{equation}%
and
\begin{eqnarray}\label{m3}
-\frac{d\varphi (z)}{dz} &=&\frac{r(z)}{(1+z)h^{2}(z)}+\frac{5\epsilon }{4}%
\frac{r^{2}(z)}{(1+z)h^{2}(z)}\nonumber\\
&& \hspace{-1.2cm} +\xi \frac{\varphi (z)}{(1+z)h^{2}(z)}
+\frac{\sigma }{2}\frac{\varphi ^{2}(z)}{(1+z)h^{2}(z)}-\frac{\varphi (z)}{%
(1+z)}.
\end{eqnarray}

 The system of differential equations (\ref{m1})-(\ref{m3}) must be solved with the initial conditions $h(0)=1$, $r(0)=r_0$, and $\varphi (0)=\varphi _0$, respectively, where  $r_0=\rho (0)/rho_c=\Omega _m$  is the density parameter of the baryonic matter. Several relations between the numerical values of the model parameters and the initial conditions  can be obtained by using Eq.~(\ref{Fr3}), from which we first obtain
\be
U_\varphi=\alpha +2\beta \varphi =6\left(\dot{H}+2H^2\right)=6H^2(1-q),
\ee
which gives the condition
\be\label{cond1}
\varphi (0)=\frac{3H_0^2}{\beta}\left(1-q(0)\right)-\frac{\alpha}{2\beta}.
\ee
For the initial value of $\psi$ we obtain the relation
\be\label{cond2}
\psi(0)=\beta \rho(0)=\beta \rho_c r(0).
\ee

\subsubsection{Model 2: $U(\varphi, \psi)=\alpha \varphi ^n \psi^{-m}+\beta \varphi$}

The second cosmological model which we will test observationally in the
scalar-tensor representation of $f(R,T)$ gravity, corresponds to a
self-interaction potential given by
\begin{equation}  \label{70}
U(\varphi, \psi)=\alpha \varphi ^n \psi^{-m}+\beta \varphi,
\end{equation}
with $\alpha$, $\beta$, $n$ and $m>0$ constants.  This form of the potential extends the de Sitter type potential (\ref{pota}) by introducing a multiplicative algebraic coupling term in its structure, which extends the $\psi$ dependence of $U(\varphi,\psi)$. The presence of this term modifies the exponential de Sitter type expansion rate, allows a time varying matter density, but still induces an accelerating expansion of the Universe. In the following we
consider again only the pressureless case, $p=0$.

By introducing the set of dimensionless variables $\left( r,h,\tau ,\alpha
_{0},\beta _{0}\right) $, defined by
\begin{equation}
H=H_{0}h, \quad \tau =H_{0}t,  \quad \rho =\rho _{c}r,  \quad  \alpha =\rho _{c}\alpha _{0}, \quad  \beta
=\rho _{c}\beta _{0},
\end{equation}%
where $\rho _{c}=3H_{0}^{2}/8\pi $ as before, we obtain the generalized Friedmann
equations in the redshift space for Model 2 as \citep{56}
\bea\label{m4}
-\frac{dh(z)}{dz}&=&\frac{n\alpha _{0}}{16\pi }\frac{\varphi ^{n-1}(z)}{%
(1+z)h(z)\psi ^{m}(z)}+\frac{\beta _{0}}{16\pi (1+z)h(z)}\nonumber\\
&&-2\frac{h(z)}{(1+z)},
\eea%
\begin{eqnarray}\label{m5}
-\frac{d\varphi (z)}{dz} &=&\frac{\beta _{0}\varphi (z)}{16\pi
(1+z)h^{2}(z)}-\frac{\varphi (z)}{(1+z)}  \nonumber\\
&&+\alpha _{0}m\frac{\varphi ^{n}(z)}{(1+z)h^{2}(z)\psi ^{m+1}(z)}\nonumber\\
&&+\frac{%
(1+3m)\alpha _{0}}{16\pi (1+z)h^{2}(z)}\frac{\varphi ^{n}(z)}{\psi ^{m}(z)},
\end{eqnarray}%
and
\begin{eqnarray}\label{m6}
-\frac{d\psi (z)}{dz} &=& \frac{\psi (z)}{\varphi (z)} \left[\frac{6h(z)(\psi (z)+8\pi
)\varphi (z)}{(1+z)h(z)\left[ (3m+1)\psi (z)+16\pi (m+1)\right] }\right. \nonumber \\
&&-\left.\frac{n(3\psi (z)+16\pi )(1+z)h(z)\frac{d\varphi (z)}{dz}}{(1+z)h(z)\left[ (3m+1)\psi (z)+16\pi (m+1)\right]}\right],
\end{eqnarray}

respectively.  The system of nonlinear differential equations (\ref{m4})-(\ref{m6}) must be solved with the initial conditions $h(0)=1$, $\varphi (0)=\varphi _0$, and $\psi (0)=\psi_0$. Once these initial conditions are fixed, the system can be solved numerically.

Note that for this system the initial values of $\varphi _{0}$
and $\psi _{0}$ are not independent.

 From the relations (\ref{Fr3}) we obtain first
\be
U_\varphi=\alpha n\varphi ^{n-1}\psi ^{-m}+\beta=6H^2\left(1-q\right),
\ee
which gives the following relation between the initial conditions and the model parameters
\be\label{45}
\alpha n\varphi_0^{n-1}\psi_0^{-m}+\beta =6H_0^2\left(1-q(0)\right).
\ee
The condition $U_\psi=-\rho$ gives the condition
\be
\alpha m \varphi ^n\psi^{-m-1}=\rho,
\ee
leading to
\be\label{cond3}
\alpha m\varphi _0^n\psi _0^{-m-1}=\rho_c r(0).
\ee

Hence, the condition (\ref{45}) can be reformulated as
\be\label{cond4}
\frac{n}{m}\rho_c\frac{\psi_0}{\varphi _0}r(0)+\beta=6H_0^2\left(1-q(0)\right).
\ee

\section{Cosmological tests }\label{cosmtests}

In the present Section, we test the observational viability of the two previously introduced cosmological models in the framework of the scalar-tensor representation of $f(R,T)$ gravity. To this aim, a detailed statistical analysis comparing the theoretical predictions of the $f(R,T)$ gravity models with the  cosmological observations is performed. In other words,  the parameters of the two models, namely, $\left(\xi,\sigma,\epsilon,h_0\right)$ for Model 1  and $\left(\alpha_0, \beta_0, \varphi,h_0\right)$ for Model 2, are constrained by  observations, using the Markov Chain Monte Carlo (MCMC) approach. To avoid any possible confusion, we mention that the parameter $h_0$ is related to the current Hubble rate as $h_0=H_{0}/100$. Furthermore,  the obtained results  are used in the cosmographic analysis of the models, i.e., by investigating the dynamical evolutions of the deceleration, the jerk, and the snap parameters, respectively. Using the  cosmographic analysis, one can figure out some general features of the behavior of the cosmological models in scalar-tensor $f(R,T)$ gravity, thus allowing for an easy  discrimination of this set of models from the $\Lambda$CDM paradigm. The observational datasets used to analyse the cosmological models in scalar-tensor $f(R,T)$ gravity are the  Hubble measurement, $H(z)$,  and the Pantheon+ data, consisting  of $57$ and $1701$ data points, respectively.

\subsection{Methodology and data description}

Nowadays, cosmological theories are tested by using different observational data, collected from different cosmological probes, such as Supernovae type Ia, Baryon Acoustic oscillations (BAO), Cosmic Microwave Background (CMB) and Hubble rate measurements, respectively. Confronting theoretical models  with  observations is the only way to distinguish between observationally supported, and unsupported models.  In order to efficiently compare theoretical models with the observational data, a detailed  statistical analysis is required, involving a multi-level approach. As cosmology is Bayesian by its very nature, parameters estimation processes are, in general, performed in the Bayesian inference framework,

\begin{equation}
\mathcal{P}(H|D)\propto \mathcal{L}(D|H),
\end{equation}

where  $\mathcal{P}(H|D)$ is the posterior distribution, and  $\mathcal{L}(D|H)$ is the likelihood function. In the case of Gaussian errors, the chi-square function, $\chi^2$, and  the likelihood function, ${\mathcal{L}}$, are related according to the relation

\begin{equation}\label{chi2_L}
\chi^2 (\theta)=-2\ln{\mathcal{L}}(\theta).
\end{equation}
In order to estimate the parameters of a given theoretical model, one has to look for the free parameters vector, $\theta$,  which minimizes the chi-square function, or, equivalently, maximizes the posterior distribution.  To constrain the cosmological parameters of the $f(R,T)$ gravity models, we minimize the $\chi^{2}$ function, by using  the Markov Chains Monte Carlo method \citep{77},

\begin{equation}
 {\chi}^2={\left( \mathcal{P}_{obs}-\mathcal{P}_{th}\right)}^{T}\hspace{0.1cm}.\hspace{0.1cm}{\bf C}^{-1}\hspace{0.1cm}.\hspace{0.1cm}\left(\mathcal{P}_{obs}-\mathcal{P}_{th}\right),
\end{equation}
where  $\mathcal{P}_{obs}$, $\mathcal{P}_{th}$, and ${\bf C}$  denote the observed values, the predicted values,  and the standard deviation, respectively.\\\\

To analyse a set of cosmological models, and in order to figure out the best model describing the observational data, a comparison criterion is usually adopted. In the current analysis, we use the most used corrected Akaike Information Criterion ($AIC_c$) \citep{87,AIC1,AIC2,AIC3,AIC4,AIC5},
\begin{equation}
AIC_{c} = \chi^2_{min}+2\mathcal{K}_{\textrm{f}}+\frac{2\mathcal{K}_{\textrm{f}}(\mathcal{K}_{\textrm{f}}+1)}{\mathcal{N}_{\textrm{t}}-\mathcal{K}_{\textrm{f}}-1},\label{AIC}
\end{equation}
where $\chi^2_{min}$, $\mathcal{K}_{\textrm{f}}$ and $\mathcal{N}_{\textrm{t}}$ denote the minimum of the chi-square function, corresponding to the best fit values,  the number of the free parameters, and the total number of the data points employed in the parameter inference process, respectively. When it comes to classify a group of cosmological models based on their statistical significance, the $AIC c$ criterion is strongly recommended.

The model with the lowest $AIC_c$ is the most supported by observations, and it is chosen to be the reference model. To measure how closely models resemble the reference one, we compute the quantity
$$
\Delta AIC_c =  AIC_{c,model}- AIC_{c,reference}.
$$
Models are said to be well supported by observations if $0 < \Delta \textrm{AIC}_c< 2$, less supported if $4 < \Delta \textrm{AIC}_c< 7$, and not supported if $\Delta \textrm{AIC}_c> 10$, with respect to the reference model.\\\\
 Similar to $AIC_c$, the Bayesian Information Criterion (BIC) \citep{BIC1,BIC2,BIC3} is also a very popular criterion used for model selection, and it is based on the Bayesian probability principles. The BIC  takes into account the goodness of fit and penalizes the complexity of models i.e. higher numbers of free parameters, stronger than the $\textrm{AIC}_c$ does. For a set of considered models, the BIC is defined as :

\begin{equation}
\textrm{BIC} = \chi^2_{min} + \mathcal{K}_{\textrm{f}} \ln(\mathcal{N}_{\textrm{t}}),
\end{equation}
where $\chi^2_{min}$, $\mathcal{K}_{\textrm{f}}$, and $\mathcal{N}_{\textrm{t}}$ were previously defined in the $\textrm{AIC}_c$ criterion. Similar to  $\textrm{AIC}$, the difference,  $\Delta$BIC, between  models and the reference one is more significant than the BIC value itself, 
\begin{equation}
\Delta \textrm{BIC} = \textrm{BIC}_{model}- \textrm{BIC}_{reference}.
\end{equation}
The model is said to be weakly disfavored by the data  if  $0<\Delta \textrm{BIC} \leq 2$, it is said to be disfavored if  $2< \textrm{BIC} \leq 6 $ while it is said to be strongly disfavored   if $\Delta \textrm{BIC} > 6$.

\subsection{Data description}

 In the following, we briefly describe the observational datasets we are going to use to test the two cosmological models introduced within the framework of the scalar-tensor representation of $f(R,T)$ gravity. To obtain the tightest possible constraints on the cosmological parameters,  probes from independent sources must be used. Given that, we estimate  the free parameters of the scalar-tensor $f(R,T)$ cosmological models by using the $H(z)$ measurements, together with the Pantheon+ data.

\subsubsection{The $H(z)$ Dataset}

In general, $H(z)$ measurements are obtained from the measurements of BAO in the radial direction of the galaxy clustering \citep{88}.  They can also be extracted from  the differential age method, using the following relation
$$
H(z)=-\frac{1}{1+z} \frac{d z}{d t}.
$$

In the current study, we use a dataset consisting of 57 points of $H(z)$,  shown in Table~\ref{T1}. The dataset consists of  31 data points derived from the differential ages of the galaxies, and from 26  points obtained from BAO. The latter are in general correlated, and one must take into account these correlations by using the corresponding covariance matrix \citep{CC}. In the literature, the 57 points are assumed most of the time to be uncorrelated, which is not a totally unacceptable approach,  if the correlations between points are negligible \citep{H57_1,H57_2,H57_3}. Therefore, the chi-square corresponding to the $H(z)$ measurements is given by
\begin{equation}
\chi_{H(z)}^{2}=\sum_{i=1}^{57} \frac{\left[H_{t h}\left(z_{i}\right)-H_{o b s}\left(z_{i}\right)\right]^{2}}{\sigma_{H\left(z_{i}\right)}^{2}},
\end{equation}
where $H_{\text {th}}$, $H_{o b s}$ and $\sigma_{H\left(z_{i}\right)}$  stand for model predictions, measured values and standard deviation, respectively.

\subsubsection{The Pantheon+ dataset}

The observations of the type Ia supernova (SNIa) have played a major role in the discovery of the comic accelerating expansion. So far, SNIa have proved to be one of the most effective tools to study the nature of the component causing the accelerating evolution of the Universe. Several  SNIa compilations have been released in recent years, like the Union \citep{106}, Union2 \citep{107} and  Union2.1 \citep{108}, Joint Light-cure Analysis (JLA) \citep{109}, Pantheon \citep{110}, and the Pantheon+ \citep{111} compilations, respectively.  The latter has been recently released, and it contains  $1701$  SNIa data, covering the redshift range $0.001<z< 2.3$. Moreover, SNIa are incredibly luminous astrophysical objects, and they are assumed to be a perfect standard candles, serving to measure relative distances via the distance modulus.

The chi-square values of the SNIa dataset are given by
\begin{equation}\label{eq_chi2}
 {\chi}_{{Pantheon+}}^2={\vec{D}}^{T}\hspace{0.1cm}.\hspace{0.1cm}{\bf C}_{{Pantheon+}}^{-1}\hspace{0.1cm}.\hspace{0.1cm}{\vec{D}},
\end{equation}
where ${\bf C}_{{Pantheon+}}$ is the covariance matrix provided with the Pantheon+ data, including both statistical and systematic uncertainties. and
\be
\vec{D}=m_{B i}-M-\mu_{\text{model}},
\ee
where $m_{B i}$ and $M$ denote the apparent, and the absolute magnitudes of SNIa, respectively. Furthermore, $\mu_{\text{model}}$ denotes the corresponding distance modulus as predicted by an assumed cosmological model, and defined as
\begin{equation}
\mu_{\text{model}}\left(z_{i}\right)=5\log_{10} \frac{D_L(z_{i})}{(H_0/c) Mpc} +25,
\end{equation}
where $H_0$ is the current value of the Hubble rate, and $D_L$ is the luminosity distance,  expressed, for a flat homogeneous and isotropic FLRW Universe,  as follows
\begin{equation}
D_L(z)=(1+z)H_0\int_{0}^{z}\frac{dz^{\prime}}{H\left(z^{\prime}\right)}.
\end{equation}

Unlike the Pantheon sample,  the degeneracy between the absolute magnitude $M$ and $H_0$ is broken in the Pantheon+ dataset, due to the rewriting of the vector $\vec{D}$, appearing in  Eq. (\ref{eq_chi2}), in terms of the distance moduli of SNIa in the Cepheid hosts,  which allows to constrain $M$ independently via the relations
\begin{equation}
\vec{D}_{i}^{\prime}= \begin{cases}m_{B i}-M-\mu_{i}^{\text {Ceph }} & i \in \text { Cepheid hosts } \\ m_{B i}-M-\mu_{\text {model }}\left(z_{i}\right) & \text { otherwise },\end{cases}
\end{equation}
where $\mu_{i}^{\rm Ceph}$ stands for the distance modulus corresponding to the Cepheid host of the $i^{\text {th }}$ SNIa,  measured independently  with Cepheid calibrators. Therefore, Eq. (\ref{eq_chi2}) can be  rewritten as

\begin{equation}\label{eq_chi2_rew}
\chi_{SN}^2 = \vec{D'}^T \cdot {\bf C}_{\text{Pantheon+}}^{-1} \cdot \vec{D'}
\end{equation}

The total chi-square function  is therefore given by
\begin{equation}
\chi^{2}_{Tot}=\chi_{H(z)}^{2}+{\chi}_{{Pantheon+}}^2.
\end{equation}

\subsection{Statistical results}

In this Section, we present the findings resulting from our analysis of the two cosmological models in the scalar-tensor representation of $f(R,T)$ gravity. We put under test the two models, Model 1 and 2, described by the differential equation systems  Eqs. (\ref{m1})-(\ref{m3}) and Eqs. (\ref{m4})-(\ref{m6}), respectively. Since these equations describing the two cosmological models do not have an exact analytical solution,   numerical solutions are required. After defining the initial conditions, by using the relations and the constraints on the initial conditions and model parameters as given by Eqs.~(\ref{cond1}), (\ref{cond2}), (\ref{cond3}) and (\ref{cond4}), respectively, and computing the numerical solutions for both models, we have constrained the free parameters by using the Hubble measurements, and the Pantheon+ data, respectively.  The priors for the cosmological parameters (Hubble function, distance modulus, matter density) have been assumed based on the previous results obtained in the framework of the $\Lambda$CDM model.


\begin{table*}
\begin{center}
{\begin{tabular}{|c|c|c|c|c|}
\hline
{\bf Model} &{\bf Parameter} &{\bf Prior} &{\bf Best fit} &{\bf Mean} \\
\hline
$  \Lambda$CDM & $\Omega_{\mathrm{m}}$ &$[0.001,1]$ &$0.279816_{-0.0108}^{+0.0108}$   &$0.279816_{-0.0108}^{+0.0108}$ \\[0.1cm]
 &$h_0$ &$[0.4,1]$ &$0.698281_{-0.0065}^{+0.0065}$       &$0.698487_{-0.0065}^{+0.0065}$ \\[0.1cm]
  &$M$   &$[-21,-17]$  &$-19.3637_{-0.0178}^{+0.0178}$   &$-19.3632_{-0.0178}^{+0.0178}$ \\[0.1cm]
\hline
\multirow{3}{*}{ $f(R,T)$ Model 1} &$\xi$     &$[0,3]$ &$1.42545_{-0.0351535}^{+0.0351535}$ &$1.430_{-0.035}^{+0.035}$ \\[0.1cm]
 &$\sigma$ &$[-3,3]$   &$-0.633484_{-0.1427}^{+0.1427}$   &$-0.80_{-0.13}^{+0.16}$ \\[0.1cm]
 &$\epsilon$ &$[-2,2]$   &$0.775255_{-0.2494}^{+0.2494}$      &$0.44_{-0.28}^{+0.24}$   \\[0.1cm]
 &$h_0$   &$[0.4,1]$  &$0.691558_{-0.0068}^{+0.0068}$   &$0.6910_{-0.0069}^{+0.0069}$ \\[0.1cm]
 &$M$   &$[-21,-17]$  &$-19.3666_{-0.0173}^{+0.0173}$   &$-19.369_{-0.018}^{+0.018}$ \\[0.1cm]
\hline
\multirow{3}{*}{ $f(R,T)$ Model 2}
 &$\alpha_0$ &$[0,3000]$   &$1323.78_{-66.4980}^{+66.4980}$   &$1306.91_{-45}^{+74}$ \\[0.1cm]
 &$\beta_0$ &$[0,100]$   &$77.7069_{-0.8321}^{+0.8321}$      &$77.74_{-0.88}^{+0.88}$   \\[0.1cm]
  &$\varphi_0$ &$[0,6]$   &$3.1181_{-0.1697}^{+0.1697}$      &$3.11913_{-0.17}^{+0.14}$   \\[0.1cm]
 &$h_0$   &$[0.4,1]$  &$0.693981_{-0.0060}^{+0.0060}$   &$0.6947_{-0.0065}^{+0.0065}$ \\[0.1cm]
 &$M$   &$[-21,-17]$  &$-19.3775_{-0.0167}^{+0.0167}$   &$-19.375_{-0.018}^{+0.018}$ \\[0.1cm]
 \hline
\end{tabular}}
\caption{ Summary of the best fit and mean values of the free cosmological parameters of the $f(R,T)$  cosmological models.}\label{tab_MCMC}
\end{center}
\end{table*}



The Table~\ref{tab_MCMC} presents  the MCMC  results regarding the  two cosmological models under consideration of the scalar-tensor  representation of $f(R,T)$ gravity. The Table~\ref{tab_MCMC} shows the prior  of the free parameters used in running the MCMC analysis, their best fit, and their mean values.  The MCMC results are then used the test the statistical significance  of each model with respect to the $\Lambda$CDM model. To this aim, we use the $AIC_c$ criterion, Eq. (\ref{AIC}), to  classify these models according to their ability of describing the cosmological data.

 According to Table~\ref{tab_AIC}, the first model of $f(R,T)$ gravity is the most favoured by observations, as it corresponds to the lowest value of  $AIC c$. As a result, it  must be considered as the reference model, i.e.  $\Delta {AIC}_c=0$. However, in the current study, we still keep   $\Lambda$CDM to be the reference model,  as it is almost always the case in the present-day literature. Therefore,  negative deviations from zero, $\Delta {AIC}_c <0$, stand for models more favored over $\Lambda$CDM, while positive deviations, $\Delta AIC {c}>0$, stand for models less favored than $\Lambda$CDM.

 In Table~\ref{tab_AIC}, we summarize the results of the statistical analysis, revealing what might be the capital result of the whole paper, i.e., {\it the fact that Model 1 is statistically better} than $\Lambda$CDM, as $\Delta AIC_{c, \textrm{Model 1}}=-5.20541$. In other words, observations favor Model 1 over the Standard Model of cosmology. On the other hand, Model 2 is  less favored than $\Lambda$CDM, but it still has considerable observational support as  $\Delta AIC_{c, \textrm{Model 2}}=1.97916$.\\\\

\begin{figure*}
\includegraphics[scale=0.6]{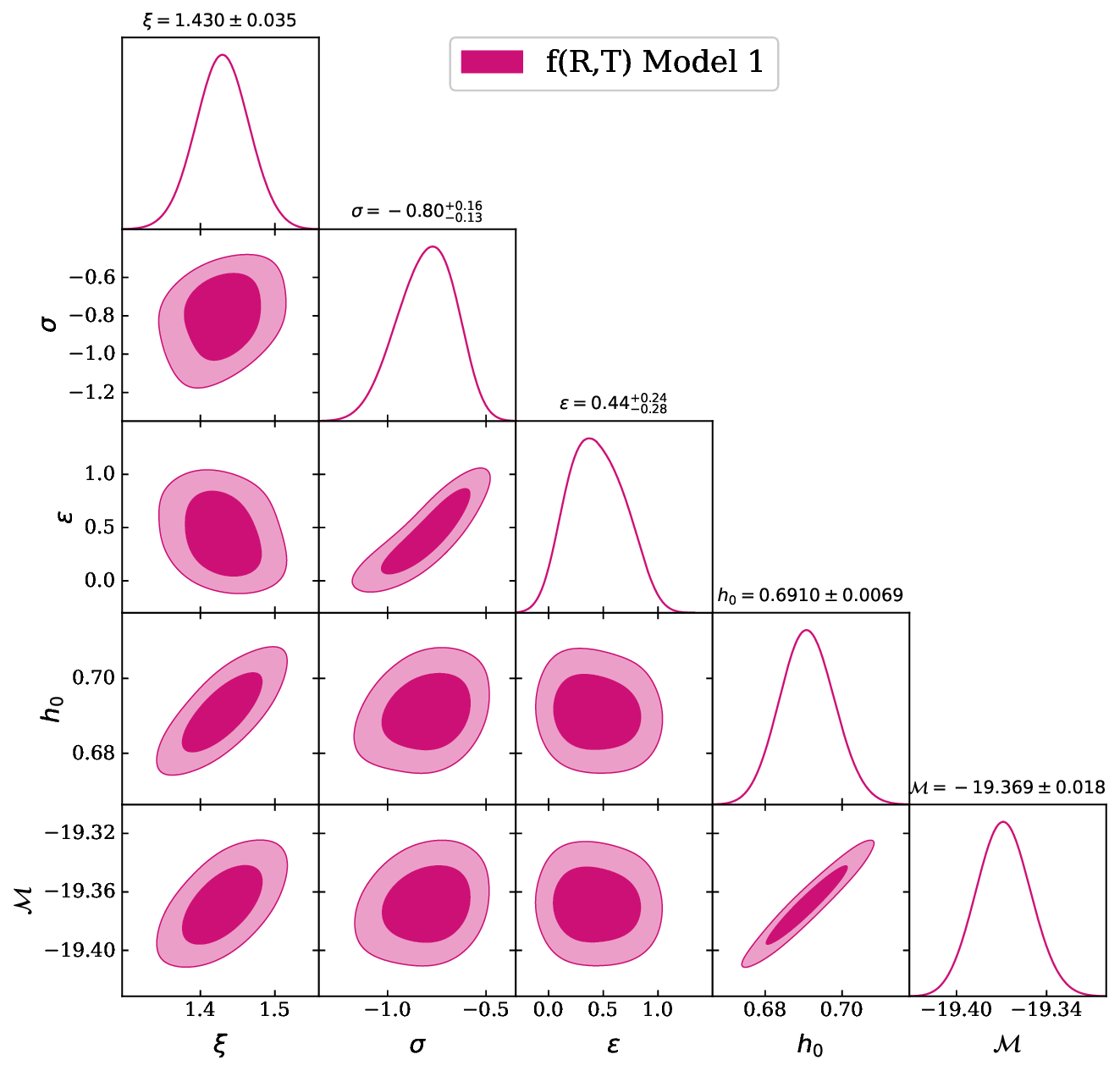}
\caption{MCMC confidence contours  at 1$\sigma$ and 2$\sigma$,  obtained after constraining the $f ( R,T )$ cosmological  Model 1 with the SNIa+$\textrm{H}(z)$ data.}\label{fig_MCMCa}
\end{figure*}

\begin{figure*}
\includegraphics[scale=0.6]{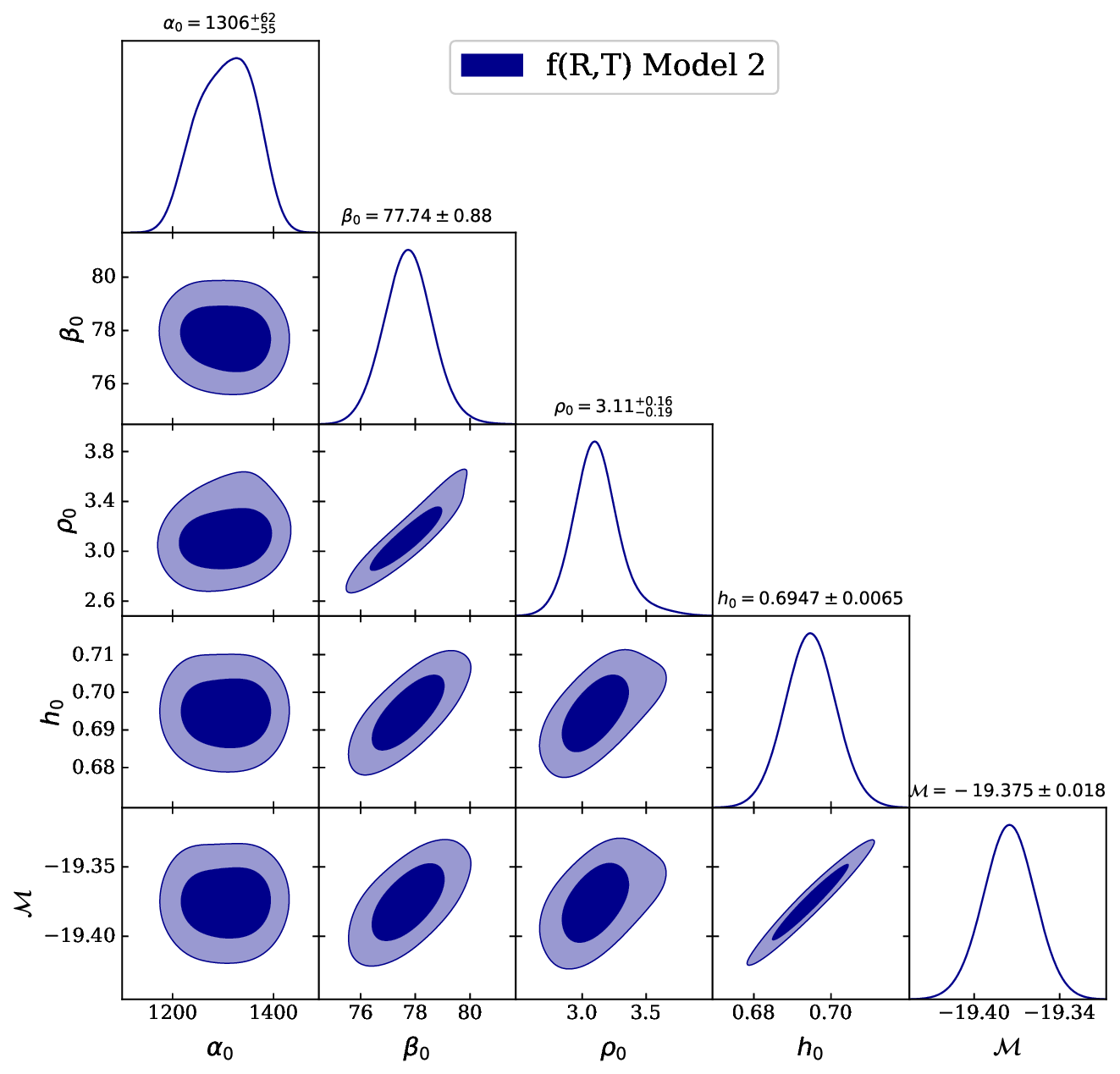}
\caption{MCMC confidence contours  at 1$\sigma$ and 2$\sigma$,  obtained after constraining the $f ( R,T )$ cosmological Model 2 with SNIa+$\textrm{H}(z)$ data.}\label{fig_MCMCb}
\end{figure*}

However, according to the BIC criterion $\Lambda$CDM  is statistically better, mainly due its simplicity as it is described with less parameters compared to  $f(R, T)$ gravity models. In addition, the BIC analysis reports less evidence against the $f(R, T)$ Model 1 than $f(R, T)$ Model 2. In other words,  $f(R, T)$ Model 1 is more favored by data and consistent with  $\Lambda$CDM than the  $f(R, T)$ Model 2.\\\\
The 1D and 2D posterior distributions, at $68.3\%$ ($1\sigma$) and $95.4\%$ ($2\sigma$) confidence levels, are depicted in Fig.~\ref{fig_MCMCa} and Fig.~\ref{fig_MCMCb} for Model 1 and Model 2, respectively. The first model presents a positive correlation between $(\xi,h)$ and  $(\sigma,h)$. A positive correlation is also seen between the $(\beta_0,h)$ and  $(\rho_0,h)$ parameters of the second model, {\it which might be an important signature to address the} $H_0$ {\it tension problem within the context of }$f(R,T)$ {\it gravity}.

\begin{figure}
\includegraphics[scale=0.43]{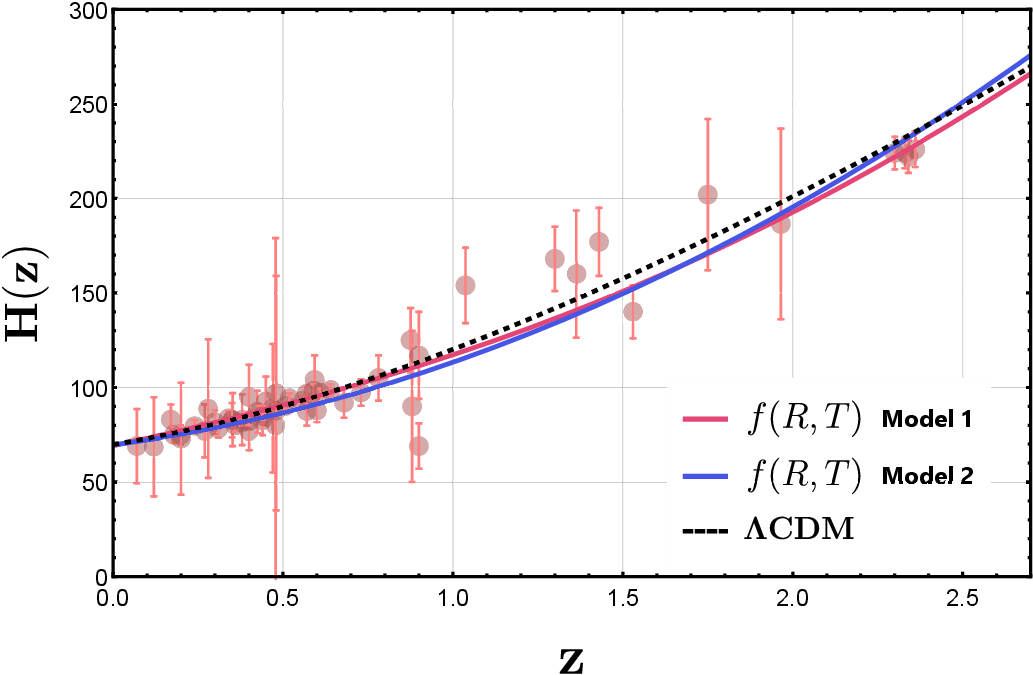}
\caption{The evolution of the  Hubble function $\mathrm{H}(\mathrm{z})$ of the $f(R,T)$ gravity and of the $\Lambda$CDM cosmological models as a function of the redshift $z$ against the Hubble data measurements. }\label{HZ}
\end{figure}

\begin{figure}
\includegraphics[scale=0.42]{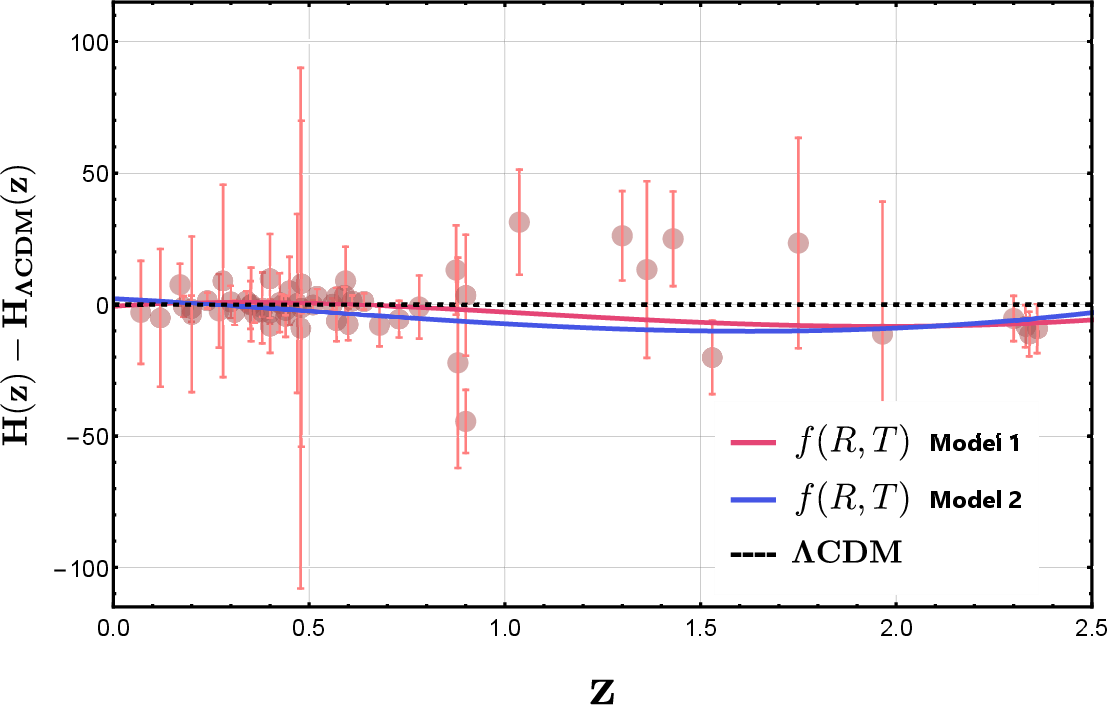}
\caption{The variation of the difference between the $f(R,T)$ cosmological models, and the $\Lambda$CDM model, as a function of the redshift $z$ against the Hubble measurements.}\label{H_diffa}
\end{figure}

Additionally, as one can see from Fig.~\ref{HZ},  both cosmological models in scalar-tensor $f(R,T)$ gravity fit well the Hubble measurement, as much as does $\Lambda$CDM,  particularly in the recent past, for low redshift values.  In Fig.~\ref{H_diffa},  the relative differences between both models and  $\Lambda$CDM are depicted. The Figure clearly shows that  Model 1 behaves exactly like the $\Lambda$CDM model up to $z=1$, while Model 2  does so only up to $z =0.5$.

\begin{table*}
\begin{center}
{\begin{tabular}{|c|c|c|c|c|c|c|c|}
\hline
{\bf Model} & ${\chi_{\text{tot}}^2}^{min} $ & $\chi_{\text {red }}^2$ &$\mathcal{K}_{\textrm{f}}$ & $A I C_c$ & $\Delta A I C_c$ &$B I C$ & {\bf $\Delta B I C$} \\[0.1cm]
\hline
$\Lambda$CDM  &$1656.4528$ &$0.943848$ &$3$  &$1662.4664$ &0 &$ 1678.87$ &  0 \\[0.1cm]
\hline
 $f(R,T)$ Model 1   &$1647.2268$   &$0.9396$   &$5$ &$1657.2610$ &$-5.20541$ &$1684.59$ & $ 5.71786$ \\[0.1cm]
\hline
 $f(R,T)$ Model 2   &$1654.4114$   &$0.94376$  &$5$  &$1664.4456$ &$1.97916$ &$ 1691.77$ & $12.9025$ \\[0.1cm]
 \hline
\end{tabular}}
\caption{Summary of ${\chi_{\text{tot}}^2}^{min} $, $\chi_{\text {red }}^2$, $A I C_c$, $\Delta A I C_c$, $B I C$ and $\Delta B I C$ for $\Lambda$CDM and $f(R,T)$  models.}\label{tab_AIC}
\end{center}
\end{table*}




\subsection{Cosmographic analysis}

In view of the existence of a considerable number of alternatives theoretical models describing the current cosmic acceleration, an approach relying on the scale factor derivatives, dubbed cosmography \citep{visser2004jerk,visser2005cosmography,cattoen2008cosmographic,visser2010cosmographic}, has proved its extreme efficiency in comparing models to each other and discriminating between the various dark energy fluid types proposed from a theoretical perspective. In other words, the cosmographic analysis offers a useful way to compare a large number of cosmological models in the space of the scale factor derivatives.

In the framework under consideration,  in the following we  consider the deceleration, the jerk and the snap parameters, defined respectively as
 \begin{align}
	q(z)&=-\frac{1}{a}\frac{d^2a}{dt^2}\left[ \frac{1}{a}\frac{da}{dt}\right] ^{-2},\\
	j(z)&= \frac{1}{a}\frac{d^3a}{dt^3}\left[ \frac{1}{a}\frac{da}{dt}\right] ^{-3},\label{j}\\
	s(z)&= \frac{1}{a}\frac{d^4a}{dt^4}\left[ \frac{1}{a}\frac{da}{dt}\right] ^{-4}.
\end{align}

\begin{figure}
\centering
\includegraphics[scale=0.45]{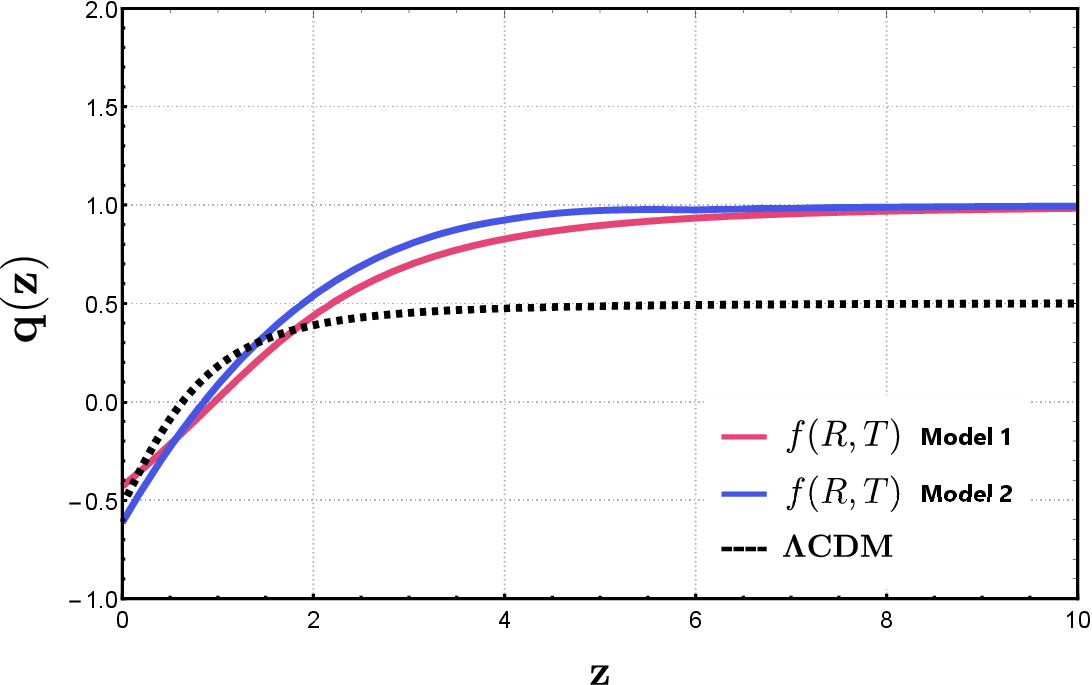}
\caption{Evolution of the deceleration parameter as a function of the redshift $z$ in the $f(R,T)$ and in $\Lambda$CDM cosmologies.}\label{q_z}
\end{figure}

\begin{figure}
\centering
\includegraphics[scale=0.45]{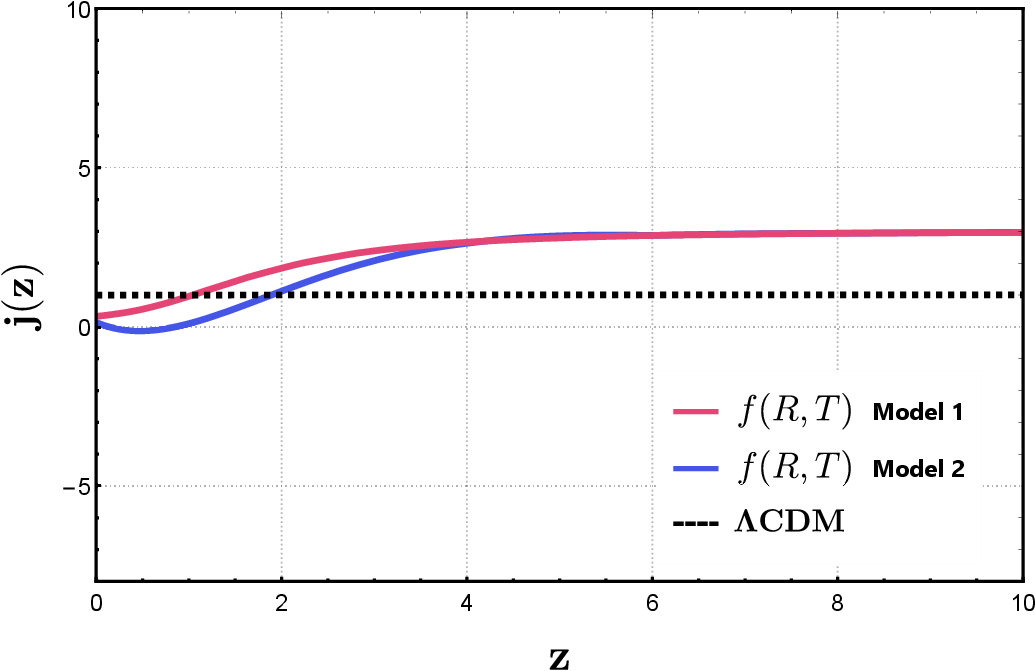}
\caption{Evolution of the jerk parameter as a function of the redshift $z$ in the $f(R,T)$ and in $\Lambda$CDM cosmological models.}\label{Jerk}
\end{figure}

\begin{figure}
\centering
\includegraphics[scale=0.45]{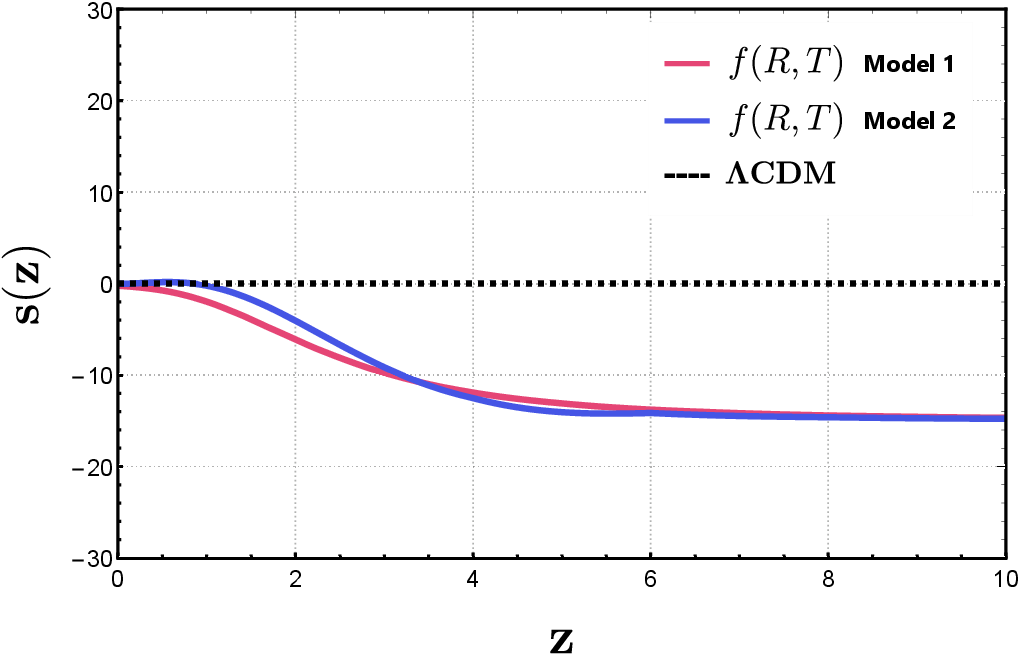}
\caption{The variation of the snap parameter $s$ as a function of the redshift $z$ in $f(R,T)$ and in  $\Lambda$CDM models.}\label{s}
\end{figure}

\begin{figure}
\includegraphics[scale=0.45]{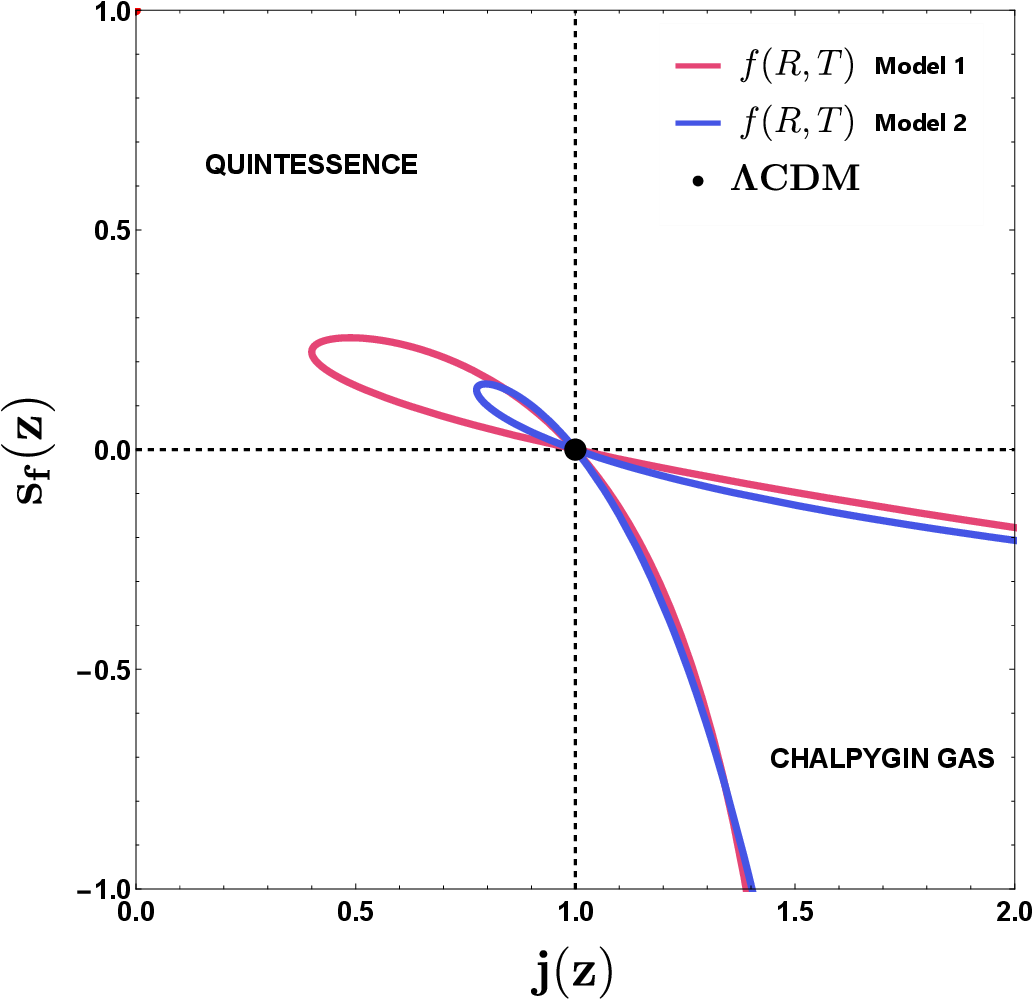}
\caption{The variation of the snap parameter $s_f$ as a function of the jerk parameter $j$ in $f(R,T)$ and in $\Lambda$CDM models.}\label{js}
\end{figure}

\begin{figure}
\centering
\includegraphics[scale=0.43]{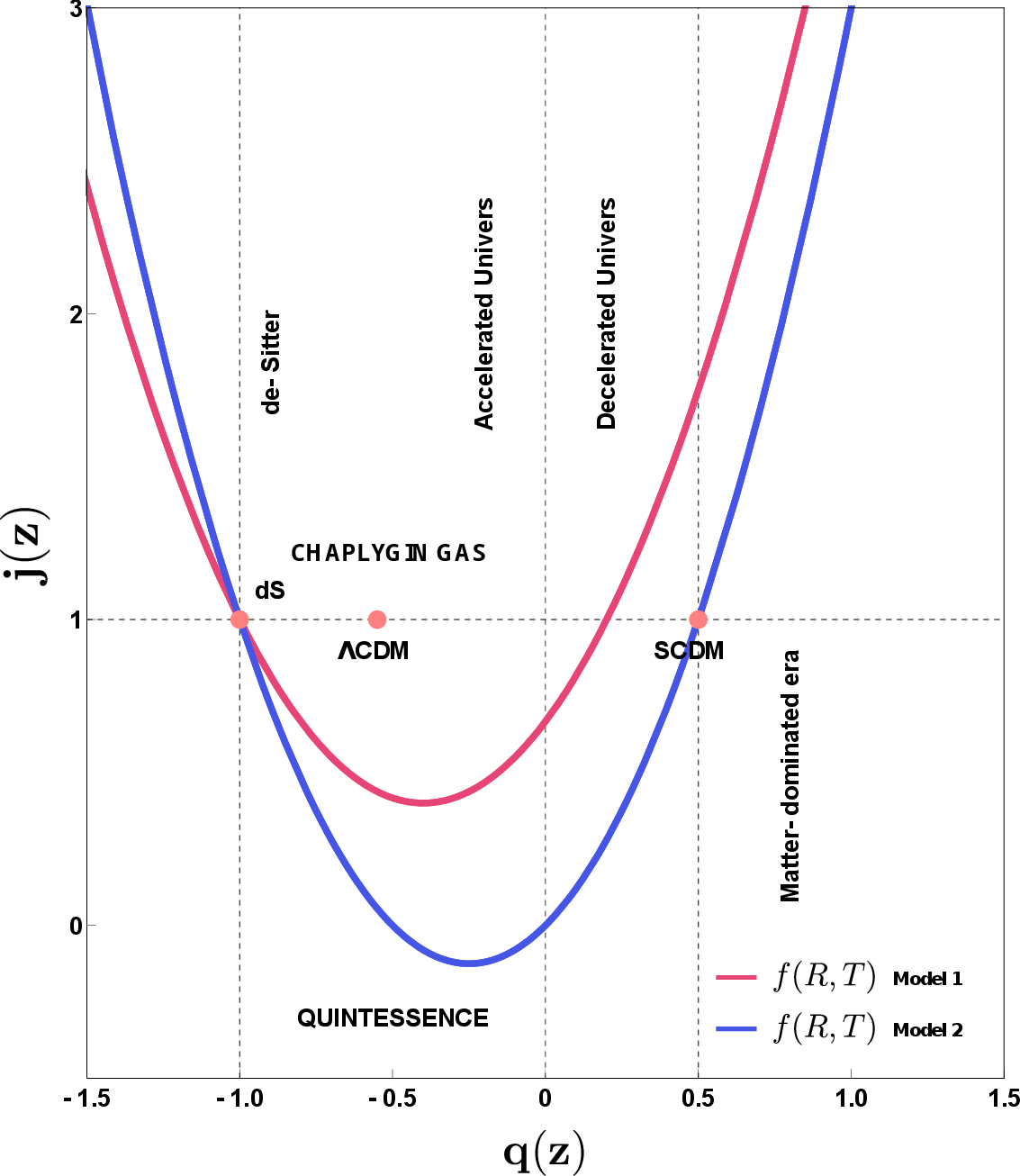}
\caption{The variation of the jerk parameter $j$  as a function of the deceleration parameter $q$ in $f(R,T)$ and $\Lambda$CDM models.}\label{qj}
\end{figure}

{\bf \em Deceleration parameter.} In Fig. \ref{q_z}, we  plot the evolution of the deceleration parameter with respect to the redshift $z$. The current values of the deceleration parameter are $-0.43$, $-0.57$ and $-0.61$ for  Model 1, $\Lambda$CDM and  Model 2, respectively. The two $f(R,T)$ models behave qualitatively as $\Lambda$CDM,  but they predict  a deceleration twice as higher as  the one  predicted by $\Lambda$CDM. Furthermore, the transition from a decelerating  to an accelerating expansion  takes place  first in Model 1, then in Model 2, and lastly  in the $\Lambda$CDM model.

{\bf \em Jerk parameter.} In Fig.~\ref{Jerk}, the evolution of the second cosmographic parameter, $j(z)$, with respect to the redshift is shown. The two analysed  $f(R,T)$ models show a similar (different) behaviour of the jerk parameter for  redshifts $z > 4$ (for $z < 4$). Furthermore,  compared to $\Lambda$CDM, both models start with  lower values  for $z<1$, before crossing the  $\Lambda$CDM curve, i.e., $j(z)=1$, and evolving toward  the constant asymptotic value  $\sim 2.5$.

{\bf \em Snap  parameter.} In Fig.~\ref{s}, the evolution of the snap parameter in terms of the redshift is depicted. Both $f(R,T)$ models show a very similar behaviour as the snap takes a negative value in the past, and reaches in the present the value $s=0$,  corresponding to  $\Lambda$CDM.

{\bf \em Statefinder diagnostic $\left\lbrace j,s_f \right\rbrace$.} The Statefinder is a purely geometrical diagnostic method, allowing  to characterize dark energy  properties in a model independent way \citep{70,alam2003exploring,69}. In addition, the  Statefinder is related to the EoS parameter and to its first time derivative, and therefore it is useful to differentiate between different dark energy types. Furthermore, the Statefinder diagnostic is also constructed with the help of the scale factor and its time derivatives.

The jerk parameter, $j$, is defined in Eq. (\ref{j}), and the parameter $s_f$ is given by the linear combination of the jerk  and the deceleration parameters  as
\begin{equation}
s_f(z)=\frac{j(z)-1}{3\left(q(z)-1/2 \right)}.
\end{equation}

To figure out the properties of the effective DE  induced by the scalar-tensor representations of the cosmological models in $f(R,T)$ gravity,  as well as to shed light on the dissimilarities between the models under study, and $\Lambda$CDM, we plot the Statefinder pair, $\left\lbrace j,s_f \right\rbrace$,  in Fig. \ref{js},  and the parametric plot of the jerk and the deceleration parameters,  $\left\lbrace q,j \right\rbrace$, in Fig. \ref{qj}. Both figures show a significant difference between the cosmological models in the scalar-tensor representations of $f(R, T)$ gravity, and  $\Lambda$CDM.\\\\

 Fig.~\ref{js} shows that both models start with values in the range $j > 1$ and $s_f < 0$ in the past, indicating the Chaplygin gas type dark energy model \citep{chap1}, and evolve to take values between $j < 1$ and $s_f > 0$, indicating the quintessence domain \citep{Quint1,Quint2} by crossing the intermediate $\Lambda$CDM fixed point $\left\lbrace j=1,s_f=0 \right\rbrace $ and then return to the Chaplygin gas domain by passing the interim fixed $\Lambda$CDM point at late times.\\\\
In Fig.~\ref{qj}, $\Lambda$CDM  separates the Chaplygin gas like (top-half) and the quintessence like (bottom half) regions. Both $f(R,T)$ models evolve from  the Chaplygin gas like to the quintessence like region. Unlike Model 1, Model 2 crosses the point $\left\lbrace j=1,q=0.5\right\rbrace $, corresponding to the Standard Cold Dark Matter cosmology (SCDM). Moreover,  the two models keep evolving in the Chaplygin gas like region, before re-entering again in the quintessence like region, crossing the steady state - the de Sitter (dS) solution,  i.e., $\left\lbrace j=1,q=-1\right\rbrace $.

\subsection{Cosmological quantities}

\begin{figure}
\centering
\includegraphics[scale=0.42]{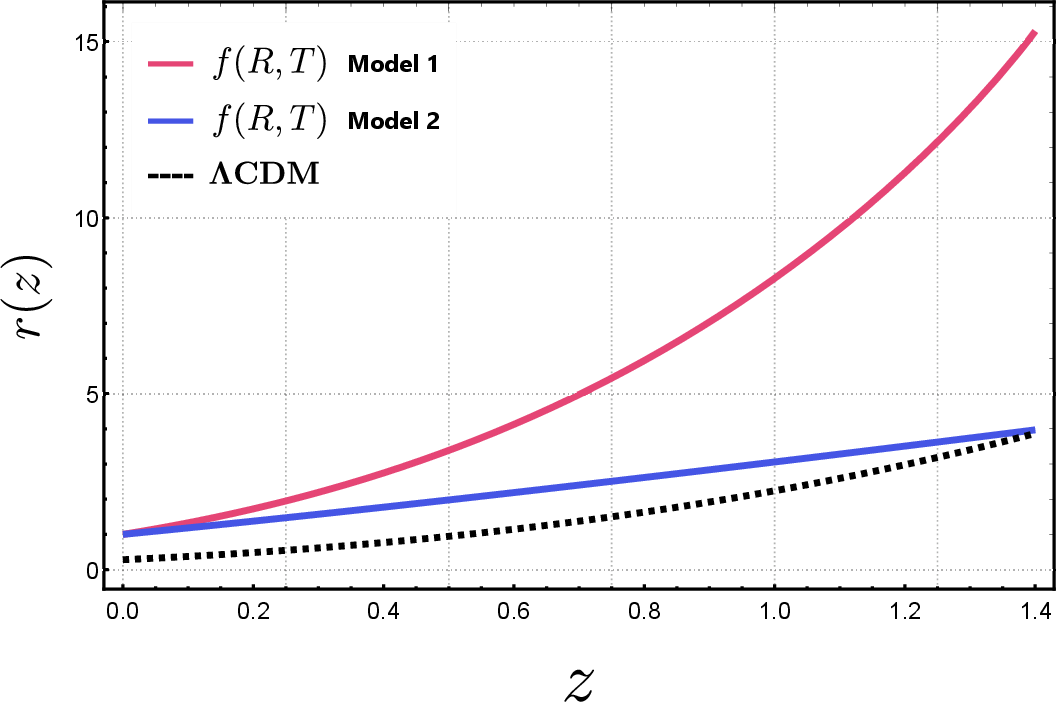}
\caption{The dynamical evolution of the energy density of matter  $r$, as a function of the redshift $z$, for both cosmological models of the scalar-tensor representation of $f(R,T)$ gravity, and for the $\Lambda$CDM model.}\label{Matter density}
\end{figure}

{ \bf \em Matter densities.} Now we proceed to a comparative study of the relevant physical parameters of the cosmological models of the $f(R,T)$ gravity and $\Lambda$CDM model. We consider first the evolution  of  the matter density parameter $r(z)$, presented in Fig.~\ref{Matter density}. The Figure shows that the ordinary matter density is a monotonically  increasing function of the redshift in all models, starting with an initial value\footnote{Other choices of  initial conditions could be adopted to compute the numerical solution.} $r(0)=1$ for $f(R,T)$ gravity models and $r(0)\approx 0.28$ for $\Lambda$CDM.

\begin{figure}
\centering
\includegraphics[scale=0.51]{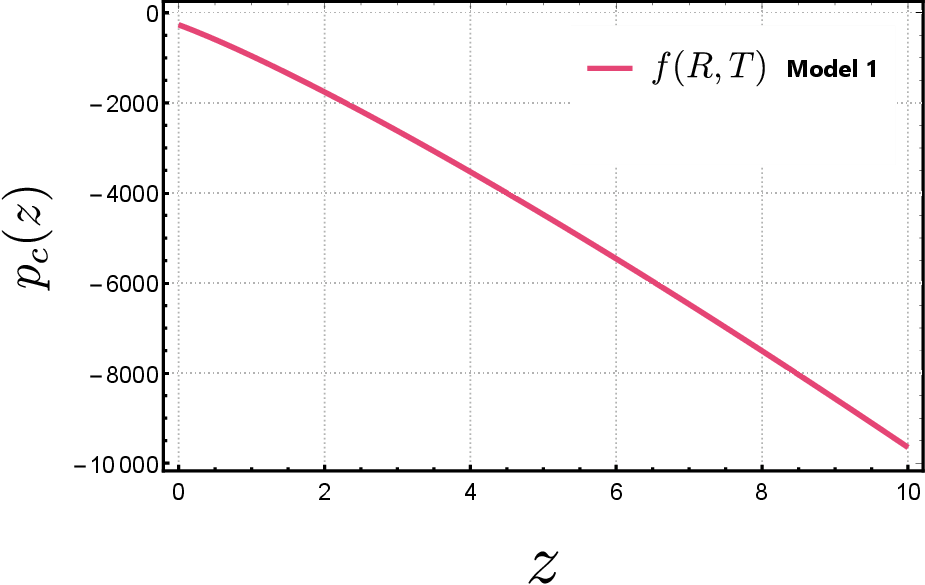}
\caption{The evolution of the creation pressure   $p_c$ as a function of redshift $z$ for Model 1. The initial conditions used for the numerical integration of the generalized Friedmann equations are $h(0)=1$, $r(0)=1$ and $\varphi(0)=-0.005$.}\label{pc_model1}
\end{figure}

\begin{figure}
\centering
\includegraphics[scale=0.45]{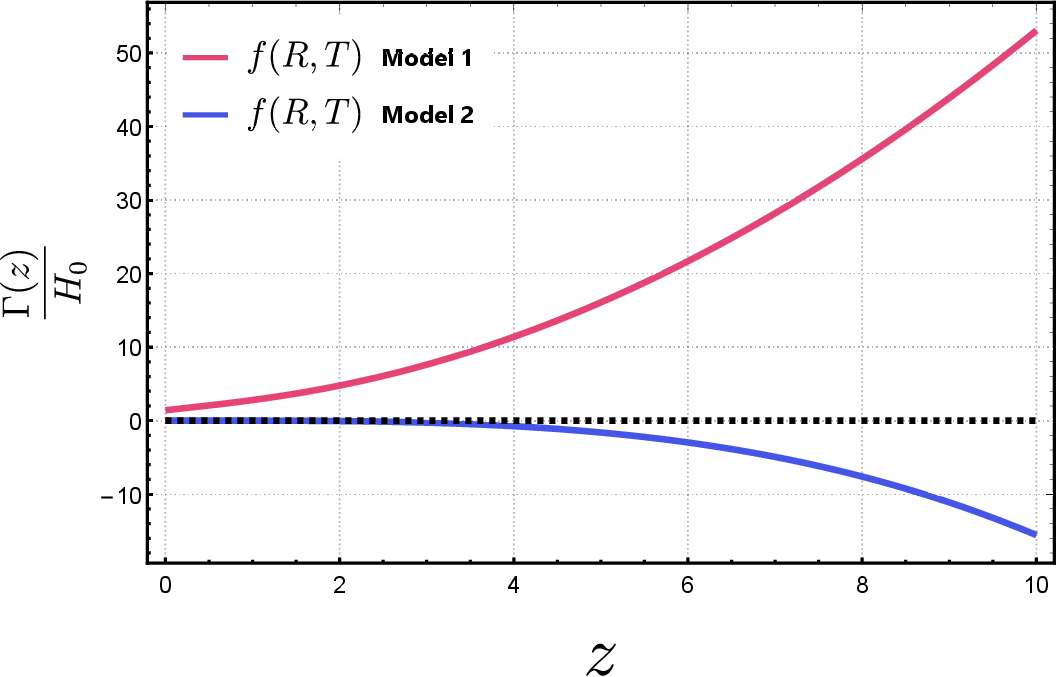}
\caption{The evolution of the matter creation rate $\Gamma/H_0$ as a function of redshift $z$. The initial conditions adopted  for the numerical solution   are $h(0)=1$, $r(0)=1$ and $\varphi(0)=-0.005$.\\
\label{creation_rate}
}
\end{figure}

 Nevertheless, at higher redshifts, important differences begin to appear between models, as the first model of $f(R,T)$ gravity  (i.e. Model 1)  predicts a higher value of matter density than Model 2, and $\Lambda$CDM. The huge amount of matter density predicted by Model 1 in the past (i.e., at higher redshifts) can be explained by  an intensive production of matter through the gravitational mechanism. The same  behavior is also shown in Fig.~\ref{pc_model1} and Fig.~\ref{creation_rate}.

 On the other hand,   a small offset resulting from the chosen initial conditions (i.e. $r(0)=1$) is separating   matter densities of the $f(R,T)$ gravity cosmological models and  $\Lambda$CDM. Nonetheless, at $z = 1.4$, this offset increases in Model 1, and disappears in Model 2, leading,  beyond that redshift,
to a similar evolution of the matter density in both Model 2 and $\Lambda$CDM.\\


\begin{figure}
\centering
\includegraphics[scale=0.45]{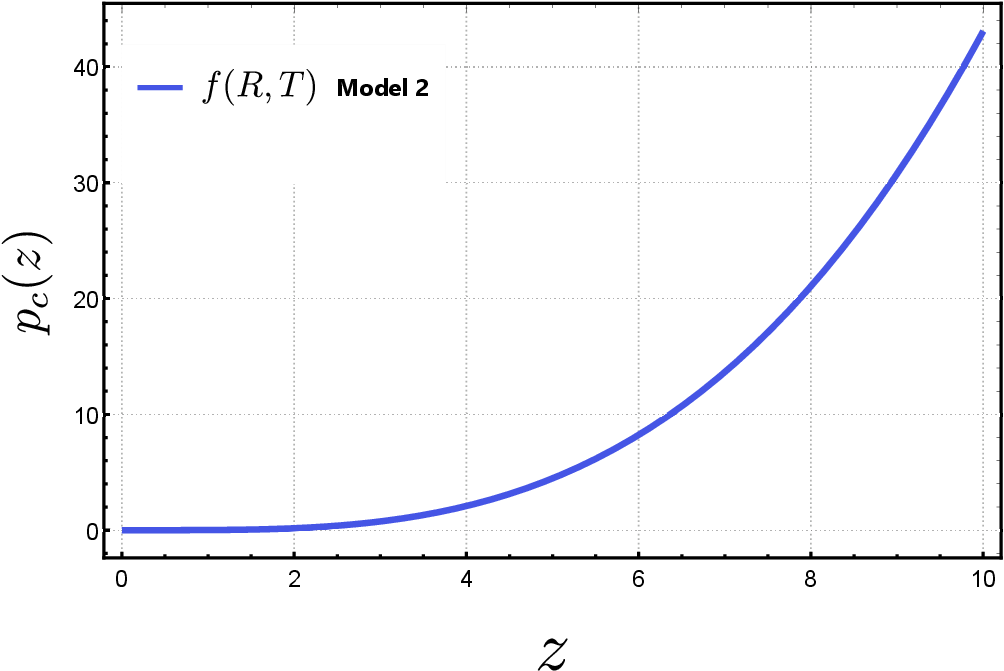}
\caption{The evolution of the creation pressure   $p_c$ as a function of redshift $z$ for Model 2. The initial conditions used for the numerical integration of the generalized Friedmann equations are $h(0)=1$, $r(0)=1$, $n=1$, $m=2$ and $\psi(0)=\left( \ m \alpha_0 \varphi_{0}^{n}\right)^{\frac{1}{1+m}} $.}\label{pc_model2}
\end{figure}


{ \bf \em Creation pressures and particle creation rates.} In the thermodynamic interpretation of $f(R,T)$ gravity, the non-conservation of the matter energy-momentum tensor is related to matter creation processes, which are described by two effective thermodynamic quantities, the creation pressure, and the particle creation rate, respectively, and which are essentially related to matter generation processes. The creation pressure for Model 1 is presented in Fig.~\ref{pc_model1}. The  creation pressure is negative in the redshift range $z\in (0,10)$, indicating a continuous presence of matter creation . As for Model 2, the creation pressure is depicted in Fig.~\ref{pc_model2}. In the late stages of the cosmological evolution, in the redshift range $z\in (0,2)$, the creation pressure is roughly a constant, and it can mimic a cosmological constant. For $z>2$ the creation pressure is positive, indicating the presence of a matter annihilation process, and increases rapidly with the redshift (decreases in time). Still, even a very small constant creation pressure can trigger the accelerated expansion of the Universe. The particle creation rates for the two $f(R,T)$ cosmological models are presented in Fig.~\ref{creation_rate}. The behavior of $\Gamma$ closely follows the behavior of the creation pressure. For Model 1, $\Gamma$ increases rapidly with the redshift, decreasing in time. Hence, the accelerating expansion of the Universe is triggered by a significant increase in the matter production rate, which reaches its maximum at the present time. On the other hand, the dynamics of the Universe for Model 2 is controlled by a very small (close to zero) particle creation/annihilation rate. For $z>4$, the particle creation rate becomes negative, indicating the beginning of a particle annihilation process. Furthermore, for redshifts higher than 4, $z > 4$, due to particle annihilation,  Model 2 violates the second law of thermodynamics. \\

{  \bf  {\em $Om(z)$ diagnostic}.}
The $Om(z)$ diagnostic tool is a very important theoretical method for distinguishing alternative cosmological models from $\Lambda$CDM. To be more precise,  $Om(z)$ diagnostic is used to determine the nature of the cosmological fluid under study, i.e., either it is a phantom fluid \citep{112,113,114,115}, a quintessence one, or a cosmological constant like.

The $Om(z)$ parameter is defined as \citep{69}
\begin{equation}\label{om}
Om(z)=\frac{\left( H(z)/H_0\right)^{2}-1}{(1+z)^{3}-1}.
\end{equation}

In the case of the $\Lambda$CDM model, the function $Om(z)$ remains  constant, and equal to the current matter density parameter $\Omega_{m0}$. Furthermore, for cosmological models with a constant EoS, i.e., $\omega={\rm constant}$, a positive slope of $Om(z)$ refers to a phantom behavior, whereas a negative slope refers to  a quintessence-like behavior. In the case of the standard  $\Lambda$CDM cosmology, and in the presence of a dynamical dark energy fluid, described by  a linear barotropic EoS, with the parameter of the equation of state denoted by $\omega (z)$, the first Friedmann equation is  written as
\bea
\left(\frac{H(z)}{H 0}\right)^{2}&=&\Omega_{m 0}(1+z)^{3}+\left(1-\Omega_{m 0}\right) \nonumber\\
&&\times \operatorname{Exp}\left[3 \int_{0}^{z} \frac{1+\omega\left(z^{\prime}\right)}{1+z^{\prime}} d z^{\prime}\right].
\eea

Thus, for a constant $\omega$, we obtain
\be
\left(\frac{H(z)}{H 0}\right)^{2}=\Omega_{m 0}(1+z)^{3}+\left(1-\Omega_{m 0}\right)(1+z)^{3(1+\omega)} .
\ee
Then, it easily follows that,
\be
Om(z)=\frac{\Omega_{m 0}(1+z)^{3}+\left(1-\Omega_{m 0}\right)(1+z)^{3(1+\omega)}-1}{(1+z)^{3}-1} .
\ee

For the $\Lambda \mathrm{CDM}$ model,  the EoS parameter is $\omega=-1$, and as a result
\be
Om(z)=\Omega_{m 0} .
\ee

\begin{figure}
\centering
\includegraphics[scale=0.45]{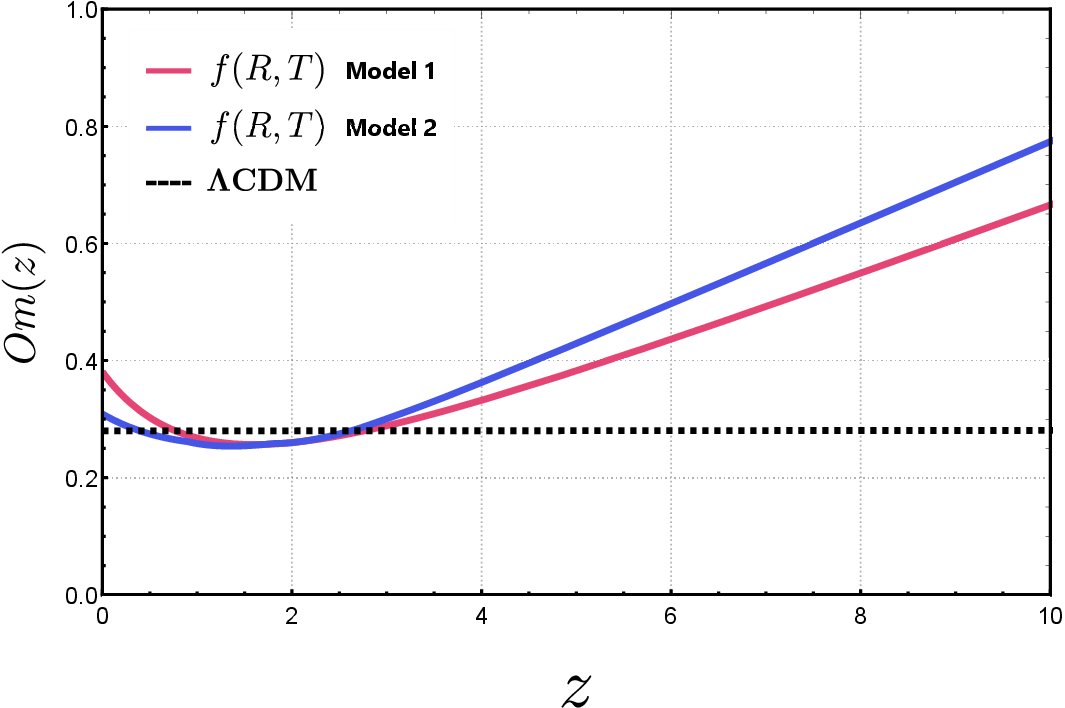}
\caption{The evolution of the $Om(z)$ function in the two $f(R,T)$ cosmological models, and in the $\Lambda$CDM model.}\label{Om_z}
\end{figure}

The evolutions  of the $Om(z)$ function with respect to the redshift in both $f(R,T)$ cosmological models, and in the $\Lambda$CDM cosmology, are presented in Fig.~\ref{Om_z}.\\

The $Om(z)$ functions for the three considered cosmological models are different. For $\Lambda$CDM,  $Om(z)$ is an absolute constant,  while in the range $z<2$, for the $f(R,T)$ cosmological models,  $Om(z)$  is a decreasing function, i.e., it has a negative slop, which indicates a quintessence like behavior. On the other hand, for $z>2$, the $Om(z)$ is an increasing function for the $f(R,T)$ models, i.e., with a positive slope, indicating the presence of a phantom like behavior.


\section{Conclusions and final remarks}\label{conclusions}

In the present paper, as a first step in our study, we have reviewed the scalar-tensor representation of $f(R,T)$ gravity, which can be reformulated in terms of two independent scalar functions $\varphi$ and $\psi$, and in the presence of a self-interaction potential $U(\varphi,\psi)$ \citep{56}. Similarly to its metric version, the energy-momentum tensor of the matter is not conserved in this formalism.   From a physical point of view, we interpret the non-conservation of the energy-momentum tensor as implying the presence of particle production/annihilation processes, which can be elegantly describe by means of the physical and mathematical approach of the irreversible thermodynamics of open systems. The cosmological evolution and the generalized Friedmann equations can also be obtained in a straightforward manner in the scalar-tensor representation, and the cosmic behavior is essentially determined by the (generally unknown) functional form of $U(\varphi,\psi)$. We have also assumed that the non-conservation of the matter energy-momentum tensor, which also appears in the scalar-tensor formulation,  implies a particle creation/annihilation, which can be described through the formalism of the irreversible thermodynamics of open systems.

To study the viability of such a mathematical setup, we have tested observationally the implications of scalar-tensor $f(R,T)$ gravity on cosmological scales. We have considered two specific cosmological models, already introduced, and investigated in  \citep{56}, denoted by $f(R,T)$ Model 1,  and $f(R,T)$ Model 2, respectively. These models correspond to two distinct choices of the self-interaction potential $U(\varphi,\psi)$ of the theory. In the first case we have adopted a simple additive quadratic structure for $U$, while the second case corresponds to a multiplicative-additive algebraic structure.

 The system of the generalized Friedmann cosmological equations considered in the present investigation of the scalar-tensor formulation of $f(R,T)$ gravity is a strongly nonlinear system of differential equations. Hence, the behavior of its solutions essentially depends on the initial conditions used for $\varphi$ and $\psi$ for the numerical integration of the system. Small changes of the initial conditions may lead to large variations in the cosmological behavior, which could not be compatible with the observational data.  However, the initial conditions for the two scalar fields are not independent, but they are determined by the initial conditions of the cosmological parameters, or, more exactly, by the initial value of the Hubble function and of the deceleration parameter. Hence, the range of variation of the initial conditions of the scalar fields is restricted by the adopted cosmological model, and they must satisfy a set of constraints for each system of differential equations describing the cosmic expansion. These conditions also allow us to significantly restrict the priors of the model parameters during the numerical fitting.

To constrain the aforementioned models observationally, we have used a set of 57 data points of  Hubble measurements,  together with the latest release of Pantheon  compilation, i.e., Pantheon+. After solving numerically the generalized Friedmann equations for the two models, the best and the mean values of the model parameters were extracted by the help of the Markov Chain Monte Carlo  analysis.  A comparative study between the $f(R,T)$ gravity cosmological models, and the $\Lambda$CDM model  has been performed by means of the corrected Aikaike Information Criterion. The results of this comparison are shown in Table~\ref{tab_AIC}. The latter Table clearly shows that the preference order of models, according to the observations is Model 1, $\Lambda$CDM, and  then Model 2. Thus, in other words, it turns out that Model 1 fits the observational data better than the $\Lambda$CDM model, indicating that $f(R,T)$ gravity holds a greater potential to explain the present-day dynamics of the Universe than standard GR.

Using the best fit parameters, we have also studied the various cosmological and physical quantities relevant for the description of the Universe. The behavior of the deceleration parameter shows that in the past, the Universe, as  described by the $f(R,T)$ cosmological models, evolves with a larger deceleration parameter than the one  predicted by the $\Lambda$CDM model. The acceleration of the Universe begins at a much higher redshift, of the order of $z\approx 4$, as compared to the redshift $z\approx 2$, at which the $\Lambda$CDM Universe begins to accelerate. The differences between the previously noted behaviors can be explained by the presence of the effective creation/annihilation processes of $f(R,T)$ gravity. Nevertheless, the two models enter in the accelerated state of expansion, with $q<0$, at almost the same redshift, $z\approx 0.5$. This dynamical evolution can be explained by considering the variation of the particle creation rate.
When the rate of creation/annihilation becomes insignificant (i.e. at late times), the  Universe enters an accelerated phase of expansion a bit earlier in both Model 1 and 2  than in $\Lambda$CDM. In other words,  the accelerated expansion  of the Universe can be fully explained by a particle production mechanism without assuming the existence of any exotic form of  dark energy. Furthermore, the statefinder and $Om(z)$ diagnostics show that the  effective fluids characterising the studied  $f(R,T)$ cosmological models in the present are quintessence like, as indicated in Fig.~\ref{js} and Fig.~\ref{Om_z}, respectively.

 An important and interesting question is the physical nature of the matter particles that are created due to the energy transfer from geometry to matter. We may conjecture that these particles contribute to, or fully represent, the dark matter content of the present-day Universe. The physical nature of the dark matter particle is not yet known, despite the intensive experimental and observational effort dedicated to finding them. One theoretically attractive possibility is that dark matter may consist of ultra-light bosonic particles, with masses in the range of $m \in \left(10^{-24}-10^{-22}\right)$ eV \citep{Witten}. From a physical point of view, such ultra-light dark matter particles created from the gravitational field could represent a pseudo Nambu-Goldstone boson. Other types  of ultra-light dark matter particle that could be generated by the gravitational field are the axions, hypothetical particles whose existence is suggested by elementary particle physics, having masses of the order of $m\leq 10^{-22 }$ eV \citep{Park}. It is important to note that very small mass particles, like axions or Nambu-Goldstone bosons, can be generated even by weak gravitational fields. Therefore, the creation of significant amounts of very low mass dark matter particles in the early Universe, from strong gravitational fields, cannot be excluded, and this process may be active in the present-day Universe, thus triggering its accelerated evolution.

Another physical possibility for the nature of the dark matter particle creation from geometry can be related to the  Scalar Field Dark Matter (SFDM) models \citep{Matos}. In these particular types of dark matter models, dark matter is considered as a real scalar field that couples minimally to gravity. The mass of the scalar field particle is assumed to have a value of the order of $m < 10^{-21}$ eV, of the same order of magnitude as the mass of the axions or Nambu-Goldstone bosons. Such particles can also be generated in the framework of the scalar-tensor representation of the $f(R,T)$ gravity, due to the natural presence of scalar fields in the theoretical formalism. The scalar field dark matter particles, as well as the other bosonic type dark matter particles,  have the important property that at very low temperatures they can form a Bose-Einstein Condensate, a special physical state in which all particles are in the same quantum ground state \citep{Boehmer, HBEC1, HBEC2}.  From a physical point of view there is a close relationship
between the Scalar Field Dark Matter and the Bose-Einstein Condensate dark matter models. In the open irreversible thermodynamic model of particle creation discussed in the present paper, the creation of the dark matter particles can also take place in the form of dynamical scalar fields, originating from the fields $\varphi$ and $\psi$. Moreover, in the present approach we assume that the two scalar fields $\varphi$ and $\psi$ do not fully decay into dark or ordinary matter. Therefore, dark matter in the form of a scalar field could be a direct result of the particle creation dynamics during the early phase in the evolution of the Universe.

The present day cosmological observations, as well as the present knowledge of the elementary particle physics cannot constrain accurately the basic predictions of the present phenomenological particle creation model. However, even without a detailed and precise knowledge of the physical processes determining the cosmological evolution, the scalar-tensor model of the modified $f(R,T)$ gravity theory allows us to make some predictions about the dynamical aspects of the Universe. In the two models we have considered the presence of the scalar fields is fully responsible for the matter content of the Universe, and for its large scale evolution.

As a conclusion and according to $\textrm{AIC}_c$ and  $\textrm{BIC}$, it can be said that at least one of the $f(R,T)$ gravity cosmological models is statistically and observationally  more supported than the $\Lambda$CDM model. Furthermore,  the presence of matter creation in these models, described by the formalism of irreversible thermodynamics of open systems, can fully explain the dynamical evolution of the Universe, and, perhaps, the formation of cosmic structure in the early Universe, provided that the particle creation rate, or, equivalently, the creation pressure, is not zero. Unlike GR, modified theories of gravity that admit non-minimal geometry-matter couplings can provide a phenomenological description of particle production, either in a matter dominated or in an empty Universe. Thus, the results presented in this work can open some new windows for the understanding of the intricate properties of the cosmic dynamics, and of the role of the matter in an evolving Universe.

\section*{Acknowledgements}

We would like to thank the anonymous referee for comments and suggestions that helped us to significantly improve our manuscript. The work of TH is supported by a grant of the Romanian Ministry of Education and Research, CNCS-UEFISCDI, project number PN-III-P4-ID-PCE-2020-2255 (PNCDI III).
FSNL acknowledges support from the Funda\c{c}\~{a}o para a Ci\^{e}ncia e a Tecnologia (FCT) Scientific Employment Stimulus contract with reference CEECINST/00032/2018, and funding from the research grant CERN/FIS-PAR/0037/2019.
FSNL and MASP acknowledge support from the Funda\c{c}\~{a}o para a Ci\^{e}ncia e a Tecnologia (FCT)  research grants UIDB/04434/2020 and UIDP/04434/2020, and through the FCT project with reference PTDC/FIS-AST/0054/2021 (``BEYond LAmbda'').
MASP also acknowledges support from the Funda\c{c}\~{a}o para a Ci\^{e}ncia e a Tecnologia (FCT)  through the Fellowship UI/BD/154479/2022.

\section*{Data Availability}
This is entirely theoretical work, and all of the results presented in
the manuscript are derived from the equations. We did not generate
any original data during the course of this study, nor did we analyse
any third-party data in this article.

\bibliographystyle{mnras}
\bibliography{mybib} 
\newpage
\section*{Appendix}
 In this Appendix, we present the Table of a total of 57 data points. Among them, 31 points are derived using the differential age technique, while the remaining 26 points are obtained through the use of BAO and other methods, all within the redshift range of $0.07 \leq z \leq 2.42$. The following Table represented in this Section describes 57 data points, which provide values of the Hubble parameter, $H(z)$, along with their corresponding errors, $\sigma_H$. Additionally, each data point is accompanied by references for further information.

\begin{table*}
\begin{center}
\begin{tabular}{|c|c|c|c|c|c|c|c|}
\multicolumn{8}{c}{\bf Observational measurements of Hubble parameter} \\
\hline $z$ & $H(z)$ & $\sigma_H$ & Ref. \hspace{0.2cm} & $z$ & $H(z)$ & $\sigma_H$ & Ref. \hspace{0.2cm} \\[0.1cm]
\hline $0.070$  & $69$ & $19.6$ & \citep{89} \hspace{0.1cm} & $1.363$ & $160$ & $33.6$ & \citep{90}\hspace{0.1cm}\\[0.1cm]
\hline $0.120$  & $68.6$ & $26.2$ & \citep{89}  \hspace{0.1cm}   & $1.965$ & $186.5$  & $50.4$    & \citep{90}\hspace{0.1cm} \\[0.1cm]
\hline $0.480$  & $97$ & $62$ & \citep{89}\hspace{0.1cm}         & $0.24$  & $79.69$  & $2.99$    & \citep{88}\hspace{0.1cm}\\[0.1cm]
\hline $0.880$  & $90$ & $40$ & \citep{89} \hspace{0.1cm}        & $0.34$  & $83.8$   & $3.66$    & \citep{88}\hspace{0.1cm}\\[0.1cm]
\hline $0.170$  & $83$ & $8$ & \citep{91}\hspace{0.1cm}          & $0.43$  & $86.45$  & $3.97$    & \citep{88}\hspace{0.1cm}\\[0.1cm]
\hline $0.270$  & $77$ & $14$ & \citep{91}\hspace{0.1cm}         & $0.30$  & $81.7$   & $6.22$    & \citep{92}\hspace{0.1cm}\\[0.1cm]
\hline $0.400$  & $95$ & $17$ & \citep{91}\hspace{0.1cm}         & $0.31$  & $78.18$  & $4.74$    & \citep{93}\hspace{0.1cm}\\[0.1cm]
\hline $0.900$  & $117$ & $23$ & \citep{91}\hspace{0.1cm}        & $0.36$  & $79.94$  & $3.38$    & \citep{93}\hspace{0.1cm}\\[0.1cm]
\hline $0.90$   & $69$ & $12$ & \citep{91}\hspace{0.1cm}          & $0.40$  & $82.04$  & $2.03$    & \citep{93}\hspace{0.1cm}\\[0.1cm]
\hline $1.300$  & $168$ & $17$ & \citep{91}\hspace{0.1cm}        & $0.44$  & $84.81$  & $1.83$    & \citep{93}\hspace{0.1cm}\\[0.1cm]
\hline $1.430$  & $177$ & $18$ & \citep{91}\hspace{0.1cm}        & $0.48$  & $87.79$  & $2.03$    & \citep{93}\hspace{0.1cm}\\[0.1cm]
\hline $1.530$  & $140$ & $14$ & \citep{91}\hspace{0.1cm}        & $0.52$  & $94.35$  & $2.64$    & \citep{93}\hspace{0.1cm}\\[0.1cm]
\hline $1.750$  & $202$ & $40$ & \citep{91}\hspace{0.1cm}        & $0.56$  & $93.34$  & $2.3$     & \citep{93}\hspace{0.1cm}\\[0.1cm]
\hline $0.1791$ & $75$ & $4$ & \citep{94}\hspace{0.1cm}       & $0.59$  & $98.48$  & $3.18$    & \citep{93}\hspace{0.1cm}\\[0.1cm]
\hline $0.1993$ & $75$ & $5$ & \citep{94}\hspace{0.1cm}       & $0.64$  & $98.82$  & $2.98$    & \citep{93}\hspace{0.1cm}\\[0.1cm]
\hline $0.3519$ & $83$ & $14$ & \citep{94}\hspace{0.1cm}      & $0.35$  & $82.7$   & $9.1$     & \citep{95}\hspace{0.1cm}\\[0.1cm]
\hline $0.593$  & $104$ & $13$ & \citep{94}\hspace{0.1cm}      & $0.38$  & $81.5$   & $1.9$     & \citep{96}\hspace{0.1cm}\\[0.1cm]
\hline $0.6797$ & $92$ & $8$ & \citep{94}\hspace{0.1cm}       & $0.51$  & $90.4$   & $1.9$     & \citep{96}\hspace{0.1cm}\\[0.1cm]
\hline $0.7812$ & $105$ & $12$ & \citep{94}\hspace{0.1cm}     & $0.61$  & $97.3$   & $2.1$     & \citep{96}\hspace{0.1cm}\\[0.1cm]
\hline $0.8754$ & $125$ & $17$ & \citep{94}\hspace{0.1cm}     & $0.44$  & $82.6$   & $7.8$     & \citep{97}\hspace{0.1cm}\\[0.1cm]
\hline $1.037$  & $154$ & $20$ & \citep{94}\hspace{0.1cm}      & $0.60$  & $87.9$   & $6.1$     & \citep{97}\hspace{0.1cm}\\[0.1cm]
\hline $0.200$  & $72.9$ & $29.6$ & \citep{98}\hspace{0.1cm}     & $0.73$  & $97.3$   & $7.0$     & \citep{97}\hspace{0.1cm}\\[0.1cm]
\hline $0.280$  & $88.8$ & $36.6$ & \citep{98}\hspace{0.1cm}     & $0.57$  & $87.6$   & $7.8$     & \citep{99}\hspace{0.1cm}\\[0.1cm]
\hline $0.3802$ & $83$ & $13.5$ & \citep{100}\hspace{0.1cm}   & $0.57$  & $96.8$   & $3.4$     & \citep{101}\hspace{0.1cm}\\[0.1cm]
\hline $0.4004$ & $77$ & $10.2$ & \citep{100}\hspace{0.1cm}   & $2.30$  & $224$    & $8.6$     & \citep{102}\hspace{0.1cm}\\[0.1cm]
\hline $0.4247$ & $87.1$ & $11.2$ & \citep{100}\hspace{0.1cm} & $2.33$  & $224$    & $8$       & \citep{103}\hspace{0.1cm}\\[0.1cm]
\hline $0.4497$ & $92.8$ & $12.9$ & \citep{100}\hspace{0.1cm}   & $2.34$  & $222$    & $8.5$   & \citep{102}\hspace{0.1cm}\\[0.1cm]
\hline $0.4783$ & $80$ & $99$ & \citep{100}\hspace{0.1cm}       & $2.36$  & $226$    & $9.3$   & \citep{104}\hspace{0.1cm}\\[0.1cm]
\hline $0.470$  & $89$ & $34$ & \citep{105}\hspace{0.1cm}   & & & &  \\
\toprule
\end{tabular}
\caption{The table shows the 57 data points of the Hubble parameter considered by the current paper.}\label{T1}
\end{center}
\end{table*}

\bsp	
\label{lastpage}
\end{document}